\documentclass[aps,prd,superscriptaddress,nofootinbib,amsmath,amsfonts,preprintnumbers,groupedaddress,showpacs,10pt,english]{revtex4-1}
\usepackage{amsmath}
\usepackage{amssymb}
\usepackage{babel}
\usepackage{wrapfig}
\usepackage{cancel}

\makeatletter

\usepackage{array,multirow,graphicx}
\usepackage{dcolumn}
\usepackage{newlfont}
\usepackage{bm}
\usepackage[colorlinks,citecolor=blue,urlcolor=blue,linkcolor=blue]{hyperref}
\usepackage[figtopcap]{subfigure}
\usepackage{color}

\begin{document}

\date{\today}

\title{New black hole solutions in three-dimensional $\mathit{f(R)}$ gravity}

\author{G.G.L. Nashed}
\email{nashed@bue.edu.eg}
\affiliation {Centre for Theoretical Physics, The British University, P.O. Box
43, El Sherouk City, Cairo 11837, Egypt}
\author{A. Sheykhi}
\email{asheykhi@shirazu.ac.ir} \affiliation{Department
of Physics, Shiraz University, Shiraz 71454, Iran\\
Biruni Observatory, Shiraz University, Shiraz 71454, Iran}

\begin{abstract}
We construct two new classes of analytical solutions in
three-dimensional spacetime and in the framework of $f(R)$
gravity. The first class represents a non-rotating black hole (BH)
while the second class corresponds to a rotating BH solution.  The
Ricci scalar of these BH solutions have non-trivial values and are
described by the gravitational mass $M$, two angular momentums $J$
and $J_1$, and an effective cosmological constant $\Lambda_{eff}$.
Moreover, these solutions do not restore the $3$-dimensional
Ba\~{n}ados-Teitelboim-Zanelli (BTZ) solutions of general
relativity (GR) which implies the novelty of the obtained BHs in
$f(R)$ gravity. Depending on the range of the parameters, these
solutions admit rotating/non-rotating asymptotically AdS/dS BH
interpretation in spite that the field equation of $f(R)$ has no
cosmological constant. Interestingly enough, we observe that in
contrast to BTZ solution which has only causal singularity and
scalar invariants are constant everywhere, the scalar invariants
of these solutions indicate strong singularity for the spacetime.
Furthermore, we construct the forms of the $f(R)$ function showing
that they behave as polynomial functions. Finally, we show that
the obtained solutions are stable from the viewpoint that heat
capacity has a positive value, and also from the condition of
Ostrogradski which state that the second derivative of $f(R)$
should have a positive value.
\end{abstract}

\pacs{04.50.Kd, 04.25.Nx, 04.40.Nr}
\keywords{$\mathbf{F(R)}$ gravitational theory, analytic spherically symmetric BHs, thermodynamics, stability, geodesic deviation.}

\maketitle
\section{\bf Introduction}
The investigations on GR solutions in $(2+1)$-dimensions provide a
powerful background to study the physical properties of gravity
and examine its viability in lower spacetime dimensions. In
particular, it can be simulated as a toy model of quantum gravity.
This property has been discovered after the arguments presented
connecting the possible links between $(2+ 1)$-dimensional
gravitation and the Chern-Simons theory
\cite{Achucarro:1986uwr,Witten:1988hc}. A great step put forwarded
by the authors of \cite{Banados:1992wn} who discovered a novel
solution of GR in $(2+1)$-dimensions. They constructed their
solution's using the SO(2, 2) gauge group with a negative
cosmological constant. This BH solution, looks like the features
of the $(3+1)$-dimensional Schwarzschild and Kerr black holes
which mean that it has a credible physical significance.

The BTZ BH solution has been given further improvements,
modifications and generalizations
\cite{Banados:1992gq,Carlip:1995qv,Banados:1998gg,Cataldo:2000qi,Ayon-Beato:2004ehj,
shey1,shey2}, and later, Witten has calculated the entropy of the
BTZ BHs \cite{Witten:2007kt}. The BTZ BH solution has been studied
by different fulfillments, from different physical viewpoints. For
example, the construction of geodesic equations of the uncharged
BTZ BHs \cite{Cruz:1994ir},  the quasi-normal modes
\cite{Cardoso:2001hn,Mikovic:2002uq,Crisostomo:2004hj,Setare:2003hm},
the scattering process of test particles
\cite{Gamboa:2000uc,Lepe:2003na}, the hydrostatic equilibrium
conditions for finite distributions \cite{Cruz:1994ar}, and
solutions for fluid distributions matching the exterior BTZ
spacetime
\cite{Garcia:2004jz,Garcia:2002rn,Bueno:2021krl,Cruz:2004tz,Gundlach:2020ovt}.
Moreover, by taking into account a non-constant coupling parameter
with the energy-scale, the scale-dependent version of the BTZ
solution has been derived and analyzed
\cite{Rincon:2018lyd,Rincon:2018dsq,Rincon:2019zxk,Fathi:2019jid,Rincon:2020izv}.
It has been also debated that, if the energy-momentum complexes of
Landau-Lifshitz and Weinberg are used for a rotating BTZ BH, the
same energy distribution is obtained \cite{Vagenas:2004zt}.

The rotating BTZ spacetime has been  generalized through the
addition of terms connected to the non-linear electrodynamics
\cite{Cataldo:2000ns,Cataldo:1999fh} and the conformal group
\cite{Oriti:2004qk}. Moreover, concerning the BH thermodynamics,
the BTZ BH solutions have been analyzed by using their critical
behavior and phase transitions. Another viewpoint of studying the
thermodynamics of the BTZ BH was carried out by calculating the
equilibrium thermodynamic fluctuations \cite{Cai:1996df}, in which
the extremal BTZ  BH with angular momentum serves as the critical
point, and the density of the states in the micro-and
grand-canonical ensembles has been calculated
\cite{Banados:1998ta}

Moreover, if the cosmological constant is dealt with as a
thermodynamical parameter, the AdS Kerr and the BTZ BH solutions
have been discussed in \cite{Wang:2006eb}. The quantum corrections
to the enthalpy and the equation of state of the uncharged BTZ BH
solutions have been investigated  \cite{Dolan:2010ha}. A general
class of BTZ BH solution has been studied from the viewpoint of
Ruppeiner geometry of the thermodynamic state space, and it is
shown that such a geometry is a flat one for the rotating BTZ and
the BTZ-Chern-Simons BHs, in the canonical ensemble
\cite{Sarkar:2006tg}.  The establishment of geometrothermodynamics
and the one introduced in \cite{Quevedo:2008ry}, is a method that
is used to specify a flat (1+1)-dimensional space of equilibrium
states, which is endowed with a thermodynamic metric.  A
generalization of geometrothermodynamics is discussed
\cite{Akbar:2011qw}, in which the thermodynamics of the charged
BTZ BH is explained in the frame of the Weinhold and Ruppeiner
geometries where it was shown that such geometry cannot describe,
the BH thermodynamics. To tackle this issue \cite{Hendi:2015rja},
a new metric (the HPEM metric) was inserted by a particular
formalism and it was shown that the corresponding Ricci scalar was
able to bring together, different types of phase transitions. The
HPEM metric was proved to give a consistent formalism to study the
thermodynamics of the BTZ BH solutions \cite{Hendi:2016pvx}. The
introduction of quantum scalar fields in the study of BH
thermodynamics was carried out   \cite{Singh:2014gva} and yields
the introduction of entanglement thermodynamics for mass-less
scalar fields. It has been also shown that the thermodynamics of
BTZ BHs can be deformed in the frame of gravity's rainbow,
however, the Gibbs free energy remains unchanged
\cite{Alsaleh:2017kzo}. The gravity of the rainbow has been
utilized to study the BH heat capacity and phase transition of BTZ
BHs \cite{Dehghani:2018qvn,Liang:2019jnj}. There are also some
other extensions of the BTZ BHs that are obtained as alternative
theories of gravity for example the Noether symmetries of the
rotating BTZ BH in $f(R)$ gravitation has been employed to create
new BTZ-type solutions \cite{Camci:2020yre}. In the same direction
some thermodynamic features of the BTZ BHs, such as the Carnot
heat engine, are investigated in the frame of massive gravity
\cite{Chougule:2018cny}. Moreover, Horndeski's action is
considered as the source field of the BTZ BH, and reduce it to the
familiar Einstein-Hilbert action involving a cosmological constant
has been studied where the usual 3-dimensional Smarr formula
by using a scaling symmetry of this reduced action
\cite{Bravo-Gaete:2014haa}. Additionally, the rotating BTZ BHs,
have been proven to display no kind of superradiance, if we
considered Dirac fields to vanish at infinity
\cite{Ortiz:2018ddt}. Some thermodynamic aspects of the rotating
BTZ BH have been investigated in \cite{Fathi:2021eig}.

The modified gravitational theory, $f(R)$,  was  introduced in the
scientific community as  an attempt to prescribe the early and
late cosmological story of the universe
\cite{DeFelice:2010aj,Bertolami:2009cd,Faraoni:2006sy,Cognola:2007zu,Zhang:2005vt,Li:2007xn,Song:2007da,Nojiri:2007cq,Nojiri:2007as,Capozziello:2018ddp,
Starobinsky:1980te}. From the recent results of the cosmological
observations, cosmological models of $f(R)$ gravity were used to
explain the transmission of deceleration and acceleration. This
yields to impose  limitations on the $f(R)$ cosmological  models
to allow for viable choices of $f(R)$ \cite{Capozziello:2014zda}.
The theories of $f(R)$ eliminate the contributions of any
curvature invariants except the Ricci scalar,  $R$ and they avoid
the Ostrogradski instability \cite{Ostrogradsky:1850fid} that us
usually exists in higher derivative gravitational theories
\cite{Woodard:2006nt}. Many BH solutions in the theories of $f(R)$
were derived and they either are deviations from the well-known BH
solutions of GR, or they have new properties that could be
discussed. Static and spherically symmetric BH solutions  in (3 +
1) and (2 + 1)-dimensions are derived and analyzed
\cite{Sebastiani:2010kv,Multamaki:2006zb,Hendi:2014wsa,Hendi:2014mba},
while charged and rotating BH solutions were explained  in
\cite{Multamaki:2006ym,Nashed:2020kdb,Nashed:2020mnp,Nashed:2019uyi,Nashed:2019tuk,Cembranos:2011sr,delaCruz-Dombriz:2012bni}.
Meanwhile static and spherically symmetric BH solutions were
discussed with constant curvature, with/without electric charge
and cosmological constant \cite{Hendi:2011eg,Eiroa:2020dip}. It is
the aim of this study, to extend the above catalog with a new
family of 3-dimensional in $f(R)$ modified theory and derive
analytic generalizations of the BTZ rotating/non-rotating metric
describing BHs. The new BH solutions display one or several
horizons and have satisfactory thermodynamical results.

The arrangement of this study is as follows: In Sec. \ref{S2} we
give the action and field equations of  ${f(R)}$ gravitational
theory. In Sec. \ref{S3} we apply the field equations of ${f(R)}$
gravity to $3$-dimensional spacetime, rotating one, having three
unknown metric potentials, $b(r)$, $b_1(r)$ and $b_2(r)$ which is
responsible for rotation. The resulting differential equations are
classified into four different cases:(i)$f_R(r)=const.$ and
$b_2(r)=0$, (ii)$f_R(r)=const.$ and $b_2(r)\neq0$,
(iii)$f_R(r)=c_0+\frac{c_1}{r^2}$ and $b_2(r)=0$, and (iv)
$f_R(r)=c_0+\frac{c_1}{r^2}$ and $b_2(r)\neq 0$ \footnote{We
assume the form of the first derivative of $f(R)$ to depend on the
radial coordinate since our present study using spherically
symmetric ansate.}. Note that here $f_R=df(R)/dR$ and we use the
chain rule i.e., $f_R=df(R)/dr\times dr/dR$. The first two cases
do not give any new results different from the BTZ BH
rotating/non-rotating of Einstein GR. The last two cases are
discussed in detail regarding their analytic solutions and their
asymptote. The most amazing thing is that our starting point of
the field equation has no cosmological constant and our derived
solutions behave as AdS/dS spacetimes. Also, we calculate the
invariants,  Kretschmann scalar, the Ricci tensor square, and the
Ricci scalar,  showing that the trace of $f(R)$  gravity on such
invariants makes the singularity stronger than those of
3-dimensional GR BHs because of the non-triviality of the Ricci
scalar associated with those solutions.  In Sec. \ref{S4} we
derive the form of $f(R)$ and its second derivative, of the last
two cases, and draw them showing that their behaviors have a
positive manner which means that our solutions avoid the
Ostrogradski instability.  In Sec. \ref{S5} we calculate the
thermodynamical quantities of the two cases (iii) and (iv) showing
their horizons, entropy, Hawking temperature, and heat capacity
analytically and graphically. In the final section, we conclude
our study with the main novel results.

\section{Field equations of $\mathit{f(R)}$ gravity}\label{S2}
In this section, we consider a three-dimensional action of
$\mathit{f(R)}$ gravity and construct the corresponding field
equations in three spacetime dimensions. It is ingredient to
mention that $\mathit{f(R)}$ gravity is an amended of Einstein
gravity and restores the general relativity in the limiting case
$\mathit{f(R)=R}$ and when $\mathit{f(R)\neq R}$ then the theory
becomes different from GR. The action of $\mathit{f(R)}$ theory
can be written as
\cite{Carroll:2003wy,1970MNRAS.150....1B,Nojiri:2003ft,Capozziello:2003gx,Capozziello:2011et,Nojiri:2010wj,Nojiri:2017ncd,Capozziello:2002rd}
\begin{eqnarray} \label{a2}
I_G=\frac{1}{2\kappa^2} \int d^3x \sqrt{-g}  \mathit{f(R)}, \  \
\end{eqnarray}
where $\kappa^2=8\pi G$, and $G$ is the Newtonian gravitational
constant and $g$ is the determinant of the metric.

Varying the above action with respect to the metric $g_{\mu \nu}$
yields the vacuum field equations of $\mathit{f(R)}$ as
\cite{2005JCAP...02..010C}
\begin{eqnarray} \label{f1}
\mathit{ R}_{\mu \nu} \mathit{f_{R}}-\frac{1}{2}g_{\mu
\nu}\mathit{f( R)}+[g_{\mu \nu}\nabla^2 -\nabla_\mu
\nabla_\nu]\mathit{ f}_{_\mathit{ R}}=0,\end{eqnarray} where
$\nabla^2\equiv\nabla_\mu \nabla^\mu$. Taking the trace of the
field equations (\ref{f1}) in 3D yields
\begin{eqnarray} \label{f3}
2\nabla^2 {\mathit f_{R}}+\mathit{ R}{f_{R}}-\frac{3 \mathit
f(R)}{2}=0 \,.
\end{eqnarray}
Using Eq. (\ref{f3}) yields the function $\mathit f(R)$  in
3-dimensional in the following form
\begin{eqnarray} \label{f3s}
\mathit f(R)=\frac{2}{3}\Big[2\nabla^2 {\mathit f_{R}}+\mathit{
R}{f_{R}}\Big]\,.\end{eqnarray}

Substituting Eq. (\ref{f3s}) in  Eq. (\ref{f1}) yields the field
equations of $\mathit f(R)$ as
\begin{eqnarray} \label{f3ss}
\mathit{ R}_{\mu \nu} \mathit{f_{R}}-\frac{1}{3}g_{\mu
\nu}\mathit{ R}\mathit{ f}_{_\mathit{ R}}+\frac{1}{3}g_{\mu
\nu}\nabla^2\mathit{ f}_{_\mathit{ R}} -\nabla_\mu
\nabla_\nu\mathit{ f}_{_\mathit{ R}}=0  \,.
\end{eqnarray}
Thus, it is important to check Eqs. (\ref{f3}) and (\ref{f3ss}) to
a spherically symmetric ansatz having two unknown functions
\cite{Karakasis:2021lnq}.
\section{The 3-dimension  black hole solutions}\label{S3}
The line element of the rotating $3$-dimensional spacetime in
the coordinates $(t,r,\phi)$ can be written as
\cite{Canate:2020btq}
\begin{eqnarray} \label{met12}
ds^2=-[b(r)-r^2b_2{}^2(r)]dt^2+2r^2b_2(r)drd\phi+\frac{dr^2}{b_1(r)}+r^2d\phi^2\,,
\end{eqnarray}
with $b(r)$, $b_1(r)$, and $b_1(r)$ are  functions depending on
the radial coordinate $r$. The Ricci scalar of the metric
(\ref{met12}) figured out as
  \begin{eqnarray} \label{Ricci}
  {\textit R(r)}=-\,\frac {2b^{2}b'_1- {
r}^{3}bb_1 b'_2{}^{2}+r   b'_1
  b b'  +2rb  { b_1} b''  -r{ b_1}
 b'_1{}^{2}+2{ b_1}b'  b}{2b^{2}r}\,,
  \end{eqnarray}
where $b\equiv b(r)$,  $b_1\equiv b_1(r)$, $b_2\equiv b_2(r)$,
$b'=\frac{db}{dr}$, $b''=\frac{d^2b}{dr^2}$,
$b'_1=\frac{db_1}{dr}$ and $b'_2=\frac{db_2}{dr}$. Plugging Eqs.
(\ref{f3}), (\ref{f3ss}) with Eq. (\ref{met12}) and by  using Eq.
(\ref{Ricci}) into the field equations (\ref{f3ss}), we get the
following non-linear differential equations, in the vacuum case
 \begin{eqnarray}\label{febtz}
&& {\mathop{\mathcal{ {\L}}}}_t{}^t=
\frac {1}{12 b^{2}r}\Bigg\{ F[b'_2(3 b
 {r}^{3}{ b_2}  b'_1-3 {r}^{3}{ b_2} { b_1}
 b' +18b {r}^{2}{ b_2}  { b_1} +4 b {r}
^{3}{ b_1}  b'_2{})-b'(2 { b_1}
 b - r{b_1}  b'+ rb'_1b) +4  b'_1 b^{2}-2 rb { b_1} b''+6 b
 {r}^{3}{ b_2} { b_1} { b''_2}]\nonumber\\
 &&+F'[4 {
b_1} b^{2}-4b  { b_1} rb'
+2 b'_1
b^{2}r +6 { b_1}b {r}^{3}{ b_2} { b'_2}] +4F''  { b_1}  b^{2}r\Bigg\}
=0\,,\nonumber\\
&&{\mathop{\mathcal{{\L}}}}_t{}^\phi=\frac {1}{4 b^{2}r}\Bigg\{2\,{ b_2}  F  rb  { b_1}b'' -2\,rF  b  { b_1}
   \left( {r}^{2}{ b_2}^{2}+b \right)  b''_2  -4\,F  b  {r}^{3}{ b_1}  { b_2}
 b'_2{}^{2}-
 \left( {r}^{2} { b_2}^{2}+b\right)  \left\{ \left( F  r{ b'_1}+2{ b_1}
   \left( 3\,F  +rF'\right)  \right) b -rF  { b_1}
  b'    \right\} { b'_2}\nonumber\\
&&  +{ b_2}
 \left[rb   \left( 2
 F' { b_1}  +F  { b'_1}
 \right) b'  -2 b^{2} \left( 2F' { b_1}  +F { b'_1}   \right) -F  r{ b_1}   b'^{2}  \right]\Bigg\}
=0\,,\nonumber\\
&&{\mathop{\mathcal{{\L}}}}_r{}^r=\frac {F r{ b_1} b'^{2}-8F'' { b_1}  b^{2}r-2F rb { b_1}
b''- \left\{ F rb'_1 -2{ b_1}  \left( 2
F +rF'  \right)
 \right\} b b' +2b
  \left[2{ b_1}  \left( b  F' +F {r}^{3}
  b'_2{}^{2}\right) - \left(F +2rF'  \right) b
 b'_1   \right] }{12r b^{2}}
=0\,,
\nonumber\\
&&{\mathop{\mathcal{{\L}}}}_\phi{}^r=\frac{ r\left[2 F b r{ b_1}
{ b''_2} +
{ b'_2}
 \left( F r { b'_1}  b - \left\{ -2b
 F'  r+F  \left(rb' -6b\right)  \right\} { b_1}  \right)
 \right]}{4b^{2}}=0\,,\nonumber\\
&&{\mathop{\mathcal{ {\L}}}}_\phi{}^\phi=\frac {1}{12r b^{2}}\Bigg\{4 F''  { b_1} b^{2}r-6F b {r}^{3}{ b_2
} { b_1} { b''_2} +4F rb {
 b_1}b'' -8F b {r}^{3}{ b_1}{ b'_2}^{2
}-3\left[  \left( F r{
 b'_1} +2{ b_1}  \left\{ 3F+rF'  \right\}
 \right) b -rF { b_1}b' \right] { b_2} {r}^{2}
{ b'_2}\nonumber\\
&& -2F r{b_1} b'^{2}+2\left[ F r{b'_1}
 +{ b_1}  \left( rF'-F   \right)  \right] b
 b' +2\, \left[\left(rF' -F  \right){ b'_1} -4F{ b_1}  \right]
 b^{2}\Bigg\}=0\,,
\label{feq}
\end{eqnarray}
where $F\equiv
F(r)=\frac{df(R(r))}{dR(r)}=\frac{df(r)}{dr}\times\frac{dr}{dR}$,
$F'=\frac{dF(r)}{dr}$, and $F''=\frac{d^2F(r)}{dr^2}$. Since we
are dealing with a spherical symmetry spacetime we assume
$f(R)=f(r)$.  Finally, the form of the trace equation given by
(\ref{f3}) takes the following form
\begin{align}
\label{trac}
&{\mathop{\mathcal{{\L}}}}=\frac {F\left[ b {r}^{3}{ b_1} b'_2{}^{2
}-2  b'_1 b
^{2}- r b'_1 b b'  -2 rb { b_1} b'' + r{ b_1} b'^{2}-2{ b_1} b'b\right]+F'\left[2b  { b_1} rb' +2
b'_1 b^{2}r+4{ b_1} b^{2}\right] +4F''{ b_1}b^{2}r -3\,f b^{
2}r}{ 2rb^{2}}=0\,.
\end{align}
Now, we are going to study special cases of the above differential
equations given by Eqs. (\ref{febtz}) and (\ref{trac}), trying to
find analytical solutions\vspace{0.2cm}:
\subsection{When  $F(r)=c_0$ and   $b_2=0$\footnote{It is important to stress that the form of $F(r)$ cannot assume  zero value in this study because when $F(r)=0$ yields that $f(R)=constant$ which is out the scope of this study.}}
When $F(r)=constant=c_0$ and $b_2=0$, the  differential equations
(\ref{feq}) reduce  to the well-known BTZ differential equation
and in that case $b_1$ and $b_2$ take the following form  \cite{Setare:2003hm}:
\begin{align}
\label{sp} b(r)=b_1(r)=\Lambda r^2-m,
\end{align}
where $\Lambda$ and $m$ are integration constants.
\subsection{When  $F(r)=c_0$ and   $b_2\neq 0$}
When  $b_2\neq 0$ and $F(r)=c_0$ the  differential equations (\ref{feq}) coincide with those presented in \cite{Sarkar:2006tg} and we get the solution of these differential equations after rescaling the constants as:
\begin{align}
\label{sp}
b(r)=b1(r)=\Lambda r^2-m+\frac{J}{r^2}\,, \qquad \qquad  b2(r)=\Lambda+\frac{\sqrt{J}}{r^2}.\end{align}
\subsection{When  $F(r)\neq constant$ and   $b_2=0$}
When  $F(r)\neq constant$, i.e., when, for example, $F(r)=c_0+\frac{c_1}{r^2}$   and when $b_2=0$ we get after rescaling the constants\footnote{There are many forms that one can assume for $F(r)$ but in this study we restrict ourselves to the form that can give reasonable physical results.}:
\begin{align}
\label{sp1}
&b(r)=C\left[r^2\,ln\Bigg(c_0+\frac{c_1}{r^2}\Bigg)-\frac{7c_1}{8c_0}-\frac{C_1 r^2}{24C_2c_0{}^3}+\frac{c_1{}^2}{4r^2c_0{}^2}-\frac{c_1{}^3}{24c_0{}^3r^4}\right]\,,\nonumber\\
 & b_1(r)=\frac{1}{\Bigg(c_0-\frac{c_1}{r^2}\Bigg)^6}\Bigg[C_1 r^2-C_2\left\{r^2c_0{}^3\,ln\Bigg(c_0+\frac{c_1}{r^2}\Bigg)-21c_1c_0{}^2+\frac{6c_1{}^2c_0}{r^2}-\frac{c_1{}^3}{r^4}\right\}\Bigg]\,.
\end{align}
The line-element of solution (\ref{sp1}) takes the form:
\begin{align}
\label{line2}
&ds^2=-\left[\frac{24c_0{}^3r^2\,ln\Bigg(c_0+\frac{c_1}{r^2}\Bigg)}{c_1}-3c_0\left[7c_0-\frac{2c_1}{r^2}\right]-\frac{c_1{}^2}{r^4}+c_2r^2\right]dt^2\nonumber\\
 &
+\frac{\Bigg(c_0-\frac{c_1}{r^2}\Bigg)^6dr^2}{\frac{24c_0{}^3r^2\,ln\Bigg(c_0+\frac{c_1}{r^2}\Bigg)}{c_1}-3c_0\left[7c_0-\frac{2c_1}{r^2}\right]-\frac{c_1{}^2}{r^4}+c_2r^2}+r^2d\phi^2\,,
\end{align}
where we have  assumed $C=\frac{24c_0{}^3}{c_1}$,  $C_1=c_2$, and $C_2=-\frac{1}{c_1}$\footnote{These assumptions of the constants $C$, $C_1$ and $C_2$ give a relations between the two unknown functions $b(r)$ and $b_1(r)$ to take the form $b1(r)=\frac{b(r)}{\Bigg(c_0-\frac{c_1}{r^2}\Bigg)^6}$.}.
The above line-element, (\ref{line2}), shows clearly that the constant $c_1$ cannot equal to zero which means that this solution can not coincides with the BTZ GR BH solutions and this yields that such BH solution is a new one. Moreover, the metric (\ref{line2})  asymptotes AdS/dS spacetime when $c_0=0$.
 Now using Eq.  (\ref{line2}) to calculate the invariants of GR we obtain the following forms:
 \begin{eqnarray} \label{inv}
&& R_{\mu \nu \rho \sigma} R^{\mu \nu \rho \sigma}= R_{\mu \nu } R^{\mu \nu }\approx C_3+\frac{C_4}{r^2}+\frac{C_5}{r^4}\,, \qquad \qquad R\approx C_{6}+\frac{C_{7}}{r^2}+\frac{C_{8}}{r^4}\,,
 \end{eqnarray}

where the constants $C_i$, $i=3\dots 8$ are  combinations of  $c_0$, $c_1$, and $c_2$, i.e.,
\begin{eqnarray}
\label{conts}
&&C_3=\frac{12}{c_0{}^{12}}\Bigg(c_2+24c_0{}^3\,lnc_0\Bigg)^2\,, \qquad C_4=\frac{4C_{3}c_1}{c_0}\,, \qquad C_{5}=-\frac{24c_1{}^2}{c_0^{14}}\Bigg(c_2+8c_0{}^3+24c_0{}^3ln c_0\Bigg)\Bigg(c_2+24c_0{}^3\,lnc_0\Bigg)\,,\nonumber\\
 &&C_{6}=-\frac{6}{c_0{}^{6}}\Bigg(c_2-24c_0{}^3\,lnc_0\Bigg)\,, \qquad C_{7}=\frac{2C_{6}c_1}{c_0}\,, \qquad C_{8}=\frac{6c_1{}^2}{c_0^{8}}\Bigg(7c_2-8c_0{}^3-168c_0{}^3ln c_0\Bigg)\,.\end{eqnarray}
Here $\Big(R_{\mu \nu \rho \sigma} R^{\mu \nu \rho \sigma}, R_{\mu
\nu} R^{\mu \nu}, R\Big)$ are the Kretschmann scalar, the Ricci
tensor square, the Ricci scalar, respectively and all  have a true
singularity at $r=0$. Moreover, the above equations show that
$c_0$ must not equal zero. It is important to stress on the fact
that  the constant $c_1$ is the main source for the deviation of
the above results from GR that has the following values
$\Big(R_{\mu \nu \rho \sigma} R^{\mu \nu \rho \sigma}, R_{\mu
\nu}R^{\mu \nu},
R\Big)=\Big(12\Lambda^2,12\Lambda^2,\mp8\Lambda\Big)$. Equation
(\ref{inv}) indicates that the leading term of the invariants
$(R_{\mu \nu \rho \sigma} R^{\mu \nu \rho \sigma},R_{\mu \nu}
R^{\mu \nu},R)$ is $(C_3,C_3,C_{6})$. Therefore, Eq. (\ref{inv})
indicates that scalar invariants of our solutions are stronger
than the BTZ spacetime of GR.
\subsection{When  $F(r)\neq constant$ and   $b_2\neq0$}
Now let us turn our attention to the general case, i.e., when $F(r)\neq constant$  $F(r)=c_0+\frac{c_1}{r^2}$ and when $b_2\neq0$ we get the following solutions:
\begin{eqnarray}
\label{sp33}
&&b(r)=\frac {{C_9}^{2}}{36{r}^{10}{c_1}^{2}C_{11}} \Bigg( 576\,C_{11}\,{c_0}^{6}{r
}^{12}\Bigg[
 \left( \ln  \left( c_0\,{r}^{2}+c_1 \right)  \right) ^{2}+4\left( \ln
 \,r  \right) ^{2}\Bigg]+c_2\,{r}^{12}+21\,C_{10}\,{r}^{10}{c_0}^{2}c_1-6\,{c_1}^{2}c_0\,{
r}^{8}C_{10}+C_{11}{c_1}^{6}\nonumber\\
 &&+24\ln  \left( c_0\,{r}^{2}+c_1
 \right)\Bigg[12\,{r}^{8}{c_0}^{4} C_{11}\,{c_1}^{2} -42\,{r}^{10}{c_0
}^{5}C_{11}\,c_1-2\,{r}^{6}{c_0}^{3} {c_1}^{3}C_{11}-\,{r}^{12}{c_0}^{3} C_{10}-96\,{r}^{12}{c_0}^{6} C_{11}\,\ln \,r\Bigg]+441\,{c_1}^{2}{c_0}^{4}{r}^{8}C_{11}\nonumber\\
 &&+48{c_0}^{3}r^6\ln
 r\Bigg[{r}^{6} C_{10}+42{c_0}^{2}{r}^{4} C_{11}c_1-12{c_0}{r}^{2}C_{11}{c_1}^{2}+2{c_1}^{3}C_{11}\Bigg]-6c_0{}^3C_{11}{r}^{2}\Bigg[
42{r}^{4}{c_0}^{3}+13{r}^{2}{c_0}^{2}{c_1}-2c_0{c_1}^{2}\Bigg]+c_0{}^3{r}^{
6}C_{10}
 \Bigg) \,,
 \nonumber\\
 &&b_1(r)=\frac {{r}^{2} }{ \left( c_0\,{r}^{2}-c_1 \right) ^{6}} \Bigg( 576\,C_{11}\,{c_0}^{6}{r}^{12} \Bigg[\left( \ln
 \left( c_0\,{r}^{2}+c_1 \right)  \right) ^{2}+4(\ln\,r)^2 \Bigg]+c_2\,{r}^{12}+21\,C_{10}\,{r}^{10}{c_0}^{2}c_1-6\,{c_1}^{2}c_0\,{
r}^{8}C_{10} \nonumber\\
 &&+
24c_0{}^3r^6\ln  \left( c_0\,{r}^{2}+c_1 \right)\Bigg[12\,{r}^{2}{c_0} C_{11}\,{c_1}^{2}-2 {c_1}^{3}C_{11}-
\,{r}^{6} C_{10}-42\,{r}^{4}{c_0}^{2} C_{11}\,c_1-
\,{r}^{6}{c_0}^{3}\ln\,r   C_{11}\Bigg]+C_{11}\,{c_1}^{6}+78\,C_{11}\,{r}^{4}{c_0}^{2}{c_1}^{4}\nonumber\\
 &&+48{c_0}^{3}r^6\ln\,r\Bigg[{r}^{6}
 C_{10}+ C_{11}\{42{c_0}^{2}{r}^{4}\,c_1-12\,{c_0}{r}^{2}{c_1}^{2}+2  {c_1}^{3}\}\Bigg]+C_{11}r^2c_0c_1{}^2\Bigg[441{c_0}^{3}{r}^{6}-
252\,{c_1}{r}^{4}{c_0}^{2}-12{c_1}^{3}\Bigg]+{c_1}^{3}{r}^{
6}C_{10}
 \Bigg)\,,\nonumber\\
 &&b_2(r)=\frac {6\,c_4\,c_1\,{r}^{6}+48\,C_9\,{c_0}^{3}{r}^{6}\ln\,r -24\,C_9\,{r}^{6}{c_0
}^{3}\ln  \left( c_0\,{r}^{2}+c_1 \right) +21\,C_9{r}^{4}{c_0}^{2}c_1-6C_9\,{r}^{2}c_0\,{c_1
}^{2}+C_9\,{c_1}^{3}}{6c_1\,{r}^{6}}\,,
\end{eqnarray}
where $c_3$, $c_4$, $C_9$, $C_{10}$ and $C_{11}$ are constants. Re-scaling  the constants by  putting $C_9=6c_1$, $C_{10}=\frac{1}{c_1}$ and $C_{11}=1$ in Eq. (\ref{sp33}) we get
\begin{eqnarray}
\label{sp3}
b(r)=&&c_2r^2+21c_0{}^2-\frac{c_0{}^3ln\Bigg(c_0+\frac{c_1}{r^2}\Bigg)}{c_1}\Bigg[24\Bigg\{1-24c_0{}^3ln\Bigg(c_0+\frac{c_1}{r^2}\Bigg)\Bigg\}r^2
+1008c_0{}^2c_1{}^2-\frac{288c_0c_1{}^3}{r^2}+\frac{48c_1{}^4}{r^4}\Bigg]\nonumber\\
 &&-\frac{3c_0c_1[2-147c_1c_0{}^3]}{r^2}+
\frac{c_1{}^2[1-252c_0{}^3c_1]}{r^4}+\frac{78c_0{}^2c_1{}^4}{r^6}-\frac{12c_0c_1{}^5}{r^8}+\frac{c_1{}^6}{r^{10}}\,,\nonumber\\
 b_1(r)=&&\frac{b(r)}{\Bigg(c_0-\frac{c_1}{r^2}\Bigg)}\,, \qquad \qquad b_2(r)={ c_4}-24c_0{}^3ln\Bigg(c_0+\frac{c_1}{r^2}\Bigg)+
\frac{21c_0{}^2c_1}{r^2}-\frac{6c_0c_1{}^2}{r^4}+\frac{c_1{}^3}{r^6}\,.
\end{eqnarray}
The line-element of the BH solution (\ref{sp3}) takes the following form:
\begin{align}
\label{line3}
&ds^2=\nonumber\\
 &-\left[\frac{ \frac{24{r}^{6}{c_0}^{3}}{c_1} \left( 1-2c_4c_1 \right)\ln \Bigg(c_0+\frac{c_1}{r^2}\Bigg) +2{c_1}^{3}c_4- \left( 1+12 c_4c_0{r}^{2} \right) {c_1}^{2}+ 6c_0{r}^{2}\left(1+ 7c_4{c_0}{r}^{2} \right) c_1 + \left(  \left( {c_4}^{2} -c_2\right) {r}^{2}-21{c_0}^{2} \right) {r}^{4} }{{r}^{4}}\right]dt^2\nonumber\\
 &
+\frac{\Bigg(c_0-\frac{c_1}{r^2}\Bigg)^6 r^4dr^2}{ \frac{24{r}^{6}{c_0}^{3}}{c_1} \left( 1-2c_4c_1 \right)\ln \Bigg(c_0+\frac{c_1}{r^2}\Bigg) +2{c_1}^{3}c_4- \left( 1+12 c_4c_0{r}^{2} \right) {c_1}^{2}+ 6c_0{r}^{2}\left(1+ 7c_4{c_0}{r}^{2} \right) c_1 + \left(  \left( {c_4}^{2} -c_2\right) {r}^{2}-21{c_0}^{2} \right) {r}^{4} }\nonumber\\
 &+r^2d\phi^2+{\frac {c_4\,{r}^{6}+48\,{c_0}^{3}{r
}^{6}\ln\,r -24\,{c_0}^{3}\ln  \left( c_0\,{r}^{2}+c_1
 \right) {r}^{6}+21\,{r}^{4}{c_0}^{2}c_1-6\,{r}^{2}c_0
\,{c_1}^{2}+{c_1}^{3}}{{r}^{4}}}\,dt\,d\phi\,.
\end{align}
 Now use Eq.  (\ref{sp3}) in order to calculate the invariants  as in the case of non-rotating case we obtain the following expressions:
 \begin{eqnarray} \label{inv1}
&& R_{\mu \nu \rho \sigma} R^{\mu \nu \rho \sigma}= R_{\mu \nu } R^{\mu \nu }\approx C_{12}+\frac{C_{13}}{r^2}+\frac{C_{14}}{r^4}\,,\qquad \qquad  R\approx C_{15}+\frac{C_{16}}{r^2}+\frac{C_{17}}{r^4}\,.
 \end{eqnarray}
 where $C_{12} \cdots$, $C_{17}$  are defined as:
 \begin{eqnarray}
\label{conts}
&&C_{12}=\frac{12}{c_0{}^{12}c_1{}^2}\Bigg(c_2c_1-24c_0{}^3\,\ln\,c_0[1-24c_0{}^3c_1\,\ln\,c_0]\Bigg)^2\,, \qquad \qquad C_{13}=\frac{4C_3c_1}{c_0}\nonumber\\
 && C_{14}=-\frac{24\Bigg(c_2c_1-24c_0{}^3\,\ln\,c_0[1-24c_0{}^3c_1\,\ln\,c_0]\Bigg)\Bigg(576\,{c_0}^{6}c_1\, \left( \ln  \left( c_0 \right)
 \right) ^{2}+ \left( -24\,{c_0}^{3}+384\,{c_0}^{6}c_1
 \right) \ln  \left( c_0 \right) +c_1\,c_2-8\,{c_0}^{3}
\Bigg)}{c_0{}^{14}}\,, \nonumber\\
 &&C_{15}=-\frac{6}{c_0{}^{6}c_1}\Bigg(c_2c_1-24c_0{}^3\,\ln\,c_0[1-24c_0{}^3c_1\,\ln\,c_0]\Bigg)\,, \qquad \qquad C_{16}=\frac{2C_0c_1}{c_0}\,, \nonumber\\
 && C_{17}=\frac{6c_0}{c_0^{8}}\Bigg(7c_2c_1-8c_0{}^3+24c_0{}^3\ln c_0[16c_0{}^3-7c_3+168c_0{}^3\ln c_0]\Bigg)\,.\end{eqnarray} Equation (\ref{inv1}) shows that all  the invariants have a true singularity at $r=0$. Moreover, the above equations show that $c_1$ must not equal zero. It is important to stress on the fact that  the constant $c_1$ is the main source for the deviation of the above results from the BH BTZ of GR.
\newpage
\section{Physical properties of the BH solutions (\ref{sp1}) and (\ref{sp3})}\label{S4}
In this section, we are going, to understand the physical properties of the BH solutions  (\ref{sp1}) and (\ref{sp3}).
\subsection{Physical properties of the non-rotating BH solution,  Eq. (\ref{sp1}) }
For the BH solution (\ref{sp1}), we  write the asymptote behaviors of the metric potentials, $b(r)$, and $b_1(r)$,  given by Eq. (\ref{sp1}) and get the following expressions:
 \begin{eqnarray}\label{mpab}
&& g_{_{_{t\,t}} }=\frac{1}{g_{_{r\,r }}}= \frac{c_2-c_0{}^3ln\Bigg(c_0+\frac{c_1}{r^2}\Bigg)}{c_1}r^2+21c_0{}^2-\frac{6c_0c_1}{r^2}-\frac{c_1{}^2}{r^4}\,,\nonumber\\
 && g_{_{t\,t }}(r\rightarrow \infty)\approx r^2\Lambda_{eff}-M+\frac{{\cal J}}{r^2}-\frac{{\cal J}_1{}^2}{r^4}-\mathcal{O}(r^{-6})\,, \nonumber\\
 &&  g_{_{t\,t }}(r\rightarrow 0)\approx  \Bigg\{\Lambda_{eff}-\frac{24c_0{}^3}{c_1}\left[\ln c_0+c_1\ln\left(\frac{r^2}{c_1}\right)\right]\Bigg\}r^2-7M+\frac{4Jc_0{}^3r^4}{c_1}-\mathcal{O}(r^{6})\,, \end{eqnarray}
where  $M=3c_0{}^2$,\,\,\, $\Lambda_{eff}=\frac{c_2c_1-24c_0{}^3ln\,c_0}{c_1}$, ${\cal J}=6c_0c_1$, ${\cal J}_1{}^2=7c_1{}^2$. Using Eq. (\ref{mpab}) in (\ref{met12}), when $b_2=0$, we get
\begin{eqnarray} \label{metaf}
& &  ds^2\approx -\Bigg[r^2\Lambda_{eff}-M+\frac{{\cal J}}{r^2}-\frac{{\cal J}_1{}^2}{r^4}\Bigg]dt^2 +
\frac{dr^2}{\Bigg[r^2\Lambda_{eff}-M+\frac{{\cal J}}{r^2}-\frac{{\cal J}_1{}^2}{r^4}\Bigg]}+r^2d\phi^2\,. \end{eqnarray}
The line element (\ref{metaf}) is asymptotically approaching AdS/dS spacetime and does not coincide with the   BTZ spacetime due to the contribution of the extra terms that come mainly from the constant $c_1$ whose source is the effect of higher-order curvature terms of  $\mathit{f(R)}$ \cite{Setare:2003hm}. Moreover, Eq. (\ref{mpab}) shows in a clear way that the constant $c_1$ cannot take the value zero which indeed indicate that the BH solution (\ref{sp1}) cannot  rerun to GR.  This means that the BH solution (\ref{sp1}) is a new  one in the $f(R)$ modified theory.

Now we are going to use Eq. (\ref{sp1})  in Eq.(\ref{Ricci})  to calculate the Ricci scalar and get:
\begin{eqnarray} \label{R1}
&&R(r)=\frac { {r}^{8}}{6
 \left( {c_0}\,{r}^{2}+{ c_1} \right) ^{2} \left( {c_0}\,{r}^{
2}-{c_1} \right) ^{7}} \Bigg(24 r^4c_0{}^3\ln  \left( { c_0}\,{r}^{2}+{c_1}
 \right)\Bigg\{{r}^{6}{{c_0}}^{3}-3\, {c_1}\,{r}^{4}{{c_0}}^{2}-9{{ c_1}}^{2}{r}^{2}{{c_0}}-5{{ c_1}}^{3}\Bigg\}-3\,{{ c_0}}^
{2}{r}^{
2}c_1[{c_2}\,{r}^{6}+12{{c_1}}^{3}]\nonumber\\
 &&+48{{
 c_0}}^{3}r^4\ln \, r\Bigg\{3\,{{c_0}}^{2}{r}^{4} { c_1}-{r}^{6}{{c_0}}^{3}+9\,{{c_0}}{r}^{2}{{c_1}}^{2}+5{{c_1}}^{3} \Bigg\} -24\,{
 c_1}\,{r}^{8}{{c_0}}^{5}+76\,{{ c_1}}^{2}{r}^{6}{{c_0}}^{4}+{{c_0}}^{3}{
c_2}\,{r}^{10}+172\,{r}^{4}{{c_0}}^{3}{{c_1}}^{3}-9\,{c_0}\,{c_2}\,{r}^{6}{{c_1}}^{2}\nonumber\\
 &&-4\,{c_0}\,{{c_1}}^{5}-5\,{c_2}\,{{ c_1}}^{3}{r}^{4} \Bigg)\,,\nonumber\\
 &&R(r\rightarrow \infty) \approx -\frac{6}{c_0{}^{6}}\Bigg(c_2-24c_0{}^3\,lnc_0\Bigg)-\frac{12c_1}{c_0{}^{7}r^2}\Bigg(c_2-24c_0{}^3\,lnc_0\Bigg)+\frac{6c_1{}^2}{r^4 c_0^{8}}\Bigg(7c_2-8c_0{}^3-168c_0{}^3ln c_0\Bigg)+\mathcal{O}(r^{-6})\,,\nonumber\\
 &&R(r\rightarrow 0) \approx \frac{24c_0r^8}{c_1{}^4}-\frac{96c_0{}^2r^{10}}{c_1{}^5}+\mathcal{O}(r^{12})\,,\nonumber\\
 &&r(R)\approx \pm2\,\sqrt {\frac {3\,\left( {c_2}c_1-24\,{{c_0}}^{3}\ln \,c_0
 \right) { c_1}}{{c_0}\,\left( 144\,{{c_0}}^{3}\ln  \,c_0-{R}\,{{c_0}}^{6}c_1-6\,{
c_2}c_1  \right) }}\,,
\qquad \qquad  \qquad \qquad r\rightarrow \infty\,.
\end{eqnarray}
 Eq. (\ref{R1}) shows that
\begin{eqnarray} \label{cond}
&& c_0>0\,, \qquad  {\textrm and}  \qquad c_1>0\,,\qquad {\textrm and} \qquad {c_2}c_1-24\,{{c_0}}^{3}\ln \,c_0>0\,, \qquad {\textrm and} \qquad 144\,{{c_0}}^{3}\ln  \,c_0-{R}\,{{c_0}}^{6}c_1-6\,{
c_2}c_1>0\,,\nonumber\\
 &&{\textrm or }\qquad   c_1<0\,,\qquad {\textrm and} \qquad {c_2}c_1-24\,{{c_0}}^{3}\ln \,c_0<0\,, \qquad {\textrm and} \qquad 144\,{{c_0}}^{3}\ln  \,c_0-{R}\,{{c_0}}^{6}c_1-6\,{
c_2}c_1>0\,,\nonumber\\
 &&{\textrm or }\qquad   c_1<0\,,\qquad {\textrm and} \qquad {c_2}c_1-24\,{{c_0}}^{3}\ln \,c_0>0\,, \qquad {\textrm and} \qquad 144\,{{c_0}}^{3}\ln  \,c_0-{R}\,{{c_0}}^{6}c_1-6\,{
c_2}c_1<0\,,\nonumber\\
 &&{\textrm or } \qquad  c_1>0\,,\qquad {\textrm and} \qquad {c_2}c_1-24\,{{c_0}}^{3}\ln \,c_0<0\,, \qquad {\textrm and} \qquad 144\,{{c_0}}^{3}\ln  \,c_0-{R}\,{{c_0}}^{6}c_1-6\,{
c_2}c_1<0\,,\end{eqnarray}
otherwise we will have an imaginary quantity. Also we stress that the constant $c_0$ must take a positive value as Eq. (\ref{R1}) shows.
   The  form of  $f(r)$,  that reproduces solution (\ref{sp1}), after using the form of $F(r)=c_0+\frac{c_1}{r^2}$,  takes the following form:
\begin{eqnarray} \label{fR1}
&&f(r)=\frac {1}{{ c_0}^{5} \left(c_0\,{r}^{2}  -c_1\right) ^{7} \left( c_0\,{r}^{2}+c_1 \right) c_1}\Bigg\{96\,{r}^{10}{c_0}^{8}c_1\, \left( c_0\,{r}^{2}+c_1 \right)  \left( 4\,c_1\,c_0 \,{r}^{2}+9\,{c_1}^{2}-{c_0}^{2}{r}^{4} \right) \ln  \left( c_0+\frac{c_1}{{r}^{2}} \right) -{c_0}^{13}{r}^{16}\nonumber\\
&& +96\,{c_0}^{3}c_1\, \left( c_0\,{r}^{2} +c_1 \right)  \left( c_0\,{r}^{2} -c_1\right) ^{7}\ln \left(\frac{ 2}{3} \right)- 14\,{c_0}^{11}{r}^{12}{c_1}^{2} + \left( 96\,{c_1}^{2}{r }^{14}+14\,{r}^{10}{c_1}^{3} \right) {c_0}^{10}-288\,{c_1}^{3}{c_0}^{9}{r}^{12}+6\,{c_0}^{12}{r}^{14}c_1\nonumber\\
&& -\left( 14\,{r}^{6}{c_1}^{5}+1088 \,{c_1}^{4}{r}^{10} \right) {c_0}^{8}+ \left( 14\,{r}^{4}{c_1}^{6}-12\,{c_1}^{2}{r}^{14}c_2-288\,{c_1}^{5}{ r}^{8} \right) {c_0}^{7}+ \left( -6\,{c_1}^{7}{r}^{2}+32\,{r }^{6}{c_1}^{6}+108\,{c_1}^{3}{r}^{12}c_2 \right) {c_0}^{6}\nonumber\\
&& + \left({c_1}^ {8} -20\,{c_1}^{4}{r}^{10}c_2 \right) {c_0}^{5}+56\,{r}^{6}{c_0}^{3}{c_1}^{6}c_2-56\,{c_1}^{7}c_2\,{r}^{4}{c_0}^{2}+24\,{c_1}^{8}c_2\,{r}^{2}c_0-4\,{c_1}^{9}c_2\Bigg\}
\,,\nonumber\\
 && f(r\rightarrow0)\approx \frac {96c_1\ln\left(\frac{2}{3c_0}\right)-c_0{}^2}{c_0{}^2c_1}-\frac{12c_1(c_2+24c_0{}^3\ln\,c_0)}{c_0{}^6r^2}
 +\frac{12c_1{}^2(3c_2+4c_0{}^3+72c_0{}^3\ln\,c_0)}{c_0{}^7r^4}+\mathcal{O}(r^{-6})\,,\nonumber\\
 && f(r\rightarrow0)\approx \frac {96c_0{}^3c_1\ln\left(\frac{2}{3}\right)-c_0{}65+4c_1c_2}{c_0{}65c_1}-\frac{32c_0}{c_1{}^3}r^6+\mathcal{O}(r^{8})\,.
\end{eqnarray}
The use of the last equation of (\ref{R1}), $r(R)$, in the first equation of (\ref{fR1}) we get:
\begin{eqnarray} \label{fR2}
&&f(R\rightarrow \infty)\approx d_1+d_2R+d_3R^2\,,\nonumber\\
 \end{eqnarray}
 where $d_i$ are constants that have the form:
 \begin{eqnarray}
&&d_1=-{\frac {1}{729}}\,{\frac {3072\,\ln c_0 c_1
\,{c_0}^{3}+729\,{c_0}^{5}-73056\,c_1\,{c_0}^{3}
\ln\,2 +69984\,c_1\,{c_0}^{3}\ln \,3+256\,c_1\,{c_0}^{3}-2788\,c_1\,c_2}{{c_0}^{5}c_1}}
\,,\nonumber\\
 &&  d_2=-{\frac {64}{6561}}\,{\frac {c_1c_0\, \left( 408\,\ln
 c_0 {c_0}^{3}-408\,{c_0}^{3}\ln
\,2+358\,{c_0}^{3}+17\,c_2 \right) }{24\,
\ln  c_0 {c_0}^{3}-c_1\,c_2}}\,,\nonumber\\
&&
d_3=-{\frac {32}{59049}}\,{\frac {{c_0}^{7}{c_1}^{2} \left( 1704
\,\ln c_0 {c_0}^{3}-1704\,{c_0}^{3}\ln
\,2 +952\,{c_0}^{3}+71\,c_2 \right) }{
 \left( 24\,\ln\,c_0{c_0}^{3}-c_1\,c_2 \right) ^{2}}}
\, .
\end{eqnarray}
To avoid the tachyonic instability, we check the Dolgov-Kawasaki
stability criterion
\cite{DeFelice:2010aj,Bertolami:2009cd,Faraoni:2006sy,Cognola:2007zu}
which states that the second derivative of the gravitational model
$f_{RR}$ must be always positive. Using the chain rule we get
\begin{eqnarray} \label{fR2}
&&f_R=\frac{df(R)}{dR}=\frac{df(r)}{dr}\frac{dr}{dR}=c_0+\frac{c_1}{r^2}\,,
\end{eqnarray}
\begin{eqnarray} \label{fR3}
&&f_{RR}=\frac{d^2f(R)}{dR^2}=\frac{df_R}{dR}=\frac{df_R(r)}{dr}\frac{dr}{dR}\nonumber\\
&&
=\left(c_0\,{r}^{2}-c_1 \right) ^{8} \left(
c_0\,{r}^{2}+c_1 \right) ^{3}c_1\Bigg\{12{r}^{10} \Bigg[ 24\,
\ln  \left(c_0\,{r}^{2}+c_1 \right) {r}^{12}{c_0}^{7}-
288\,\ln  \left(c_0\,{r}^{2}+c_1 \right)c_1\,{r}^{10}
{c_0}^{6}-1008\,\ln  \left(c_0\,{r}^{2}+c_1 \right) {c_1}^{2}{r}^{8}{c_0}^{5}\nonumber\\
&&-1056\,\ln  \left(c_0\,{r}^{2}+
c_1 \right) {c_1}^{3}{r}^{6}{c_0}^{4}-360\,\ln  \left(
c_0\,{r}^{2}+c_1 \right) {c_1}^{4}{r}^{4}{c_0}^{3}
-48\,\ln \,r {r}^{12}{c_0}^{7}+576\,{c_0}^{6}{
r}^{10}\ln\,rc_1-32\,c_1\,{r}^{10}{c_0}
^{6}\nonumber\\
&&+2016\,{c_0}^{5}{r}^{8}{c_1}^{2}\ln\,r +
280\,{c_1}^{2}{r}^{8}{c_0}^{5}+2112\,{c_0}^{4}{r}^{6}{c_1}^{3}\ln \,r+816\,{c_1}^{3}{r}^{6}{c_0
}^{4}+656\,{c_1}^{4}{r}^{4}{c_0}^{3}+720\,{c_0}^{3}{r}^
{4}{c_1}^{4}\ln\, +80\,{c_1}^{5}{r}^{2}{c_0}^{2}\nonumber\\
&&-8\,{c_1}^{6}c_0-c_2\,{r}^{12}{c_0}^
{4}c_1+12\,c_2\,{c_1}^{2}{r}^{10}{c_0}^{3}+42\,{
c_1}^{3}c_2\,{r}^{8}{c_0}^{2}+44\,{c_1}^{4}c_2\,{r}^{6}c_0+15\,{c_1}^{5}c_2\,{r}^{4} \Bigg] \Bigg\}^{-1}\,.
\end{eqnarray}

\begin{figure}
\centering
\subfigure[~The behavior of the Ricci scalar given by Eq. (\ref{R1})]{\label{fig:R}\includegraphics[scale=0.33]{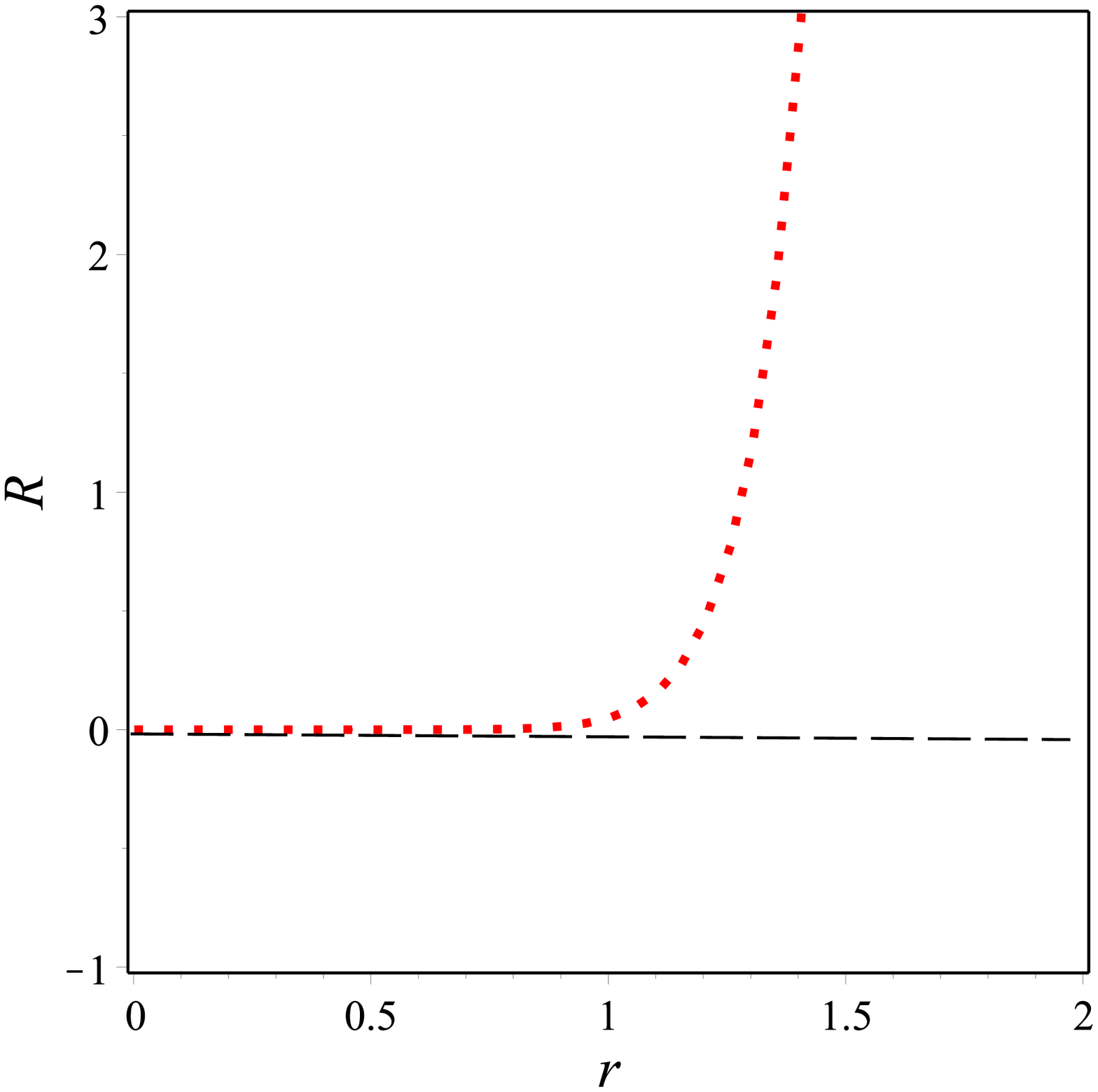}}
\subfigure[~The behavior of the function $f(r)$  given   (\ref{fR1})]{\label{fig:fr}\includegraphics[scale=0.33]{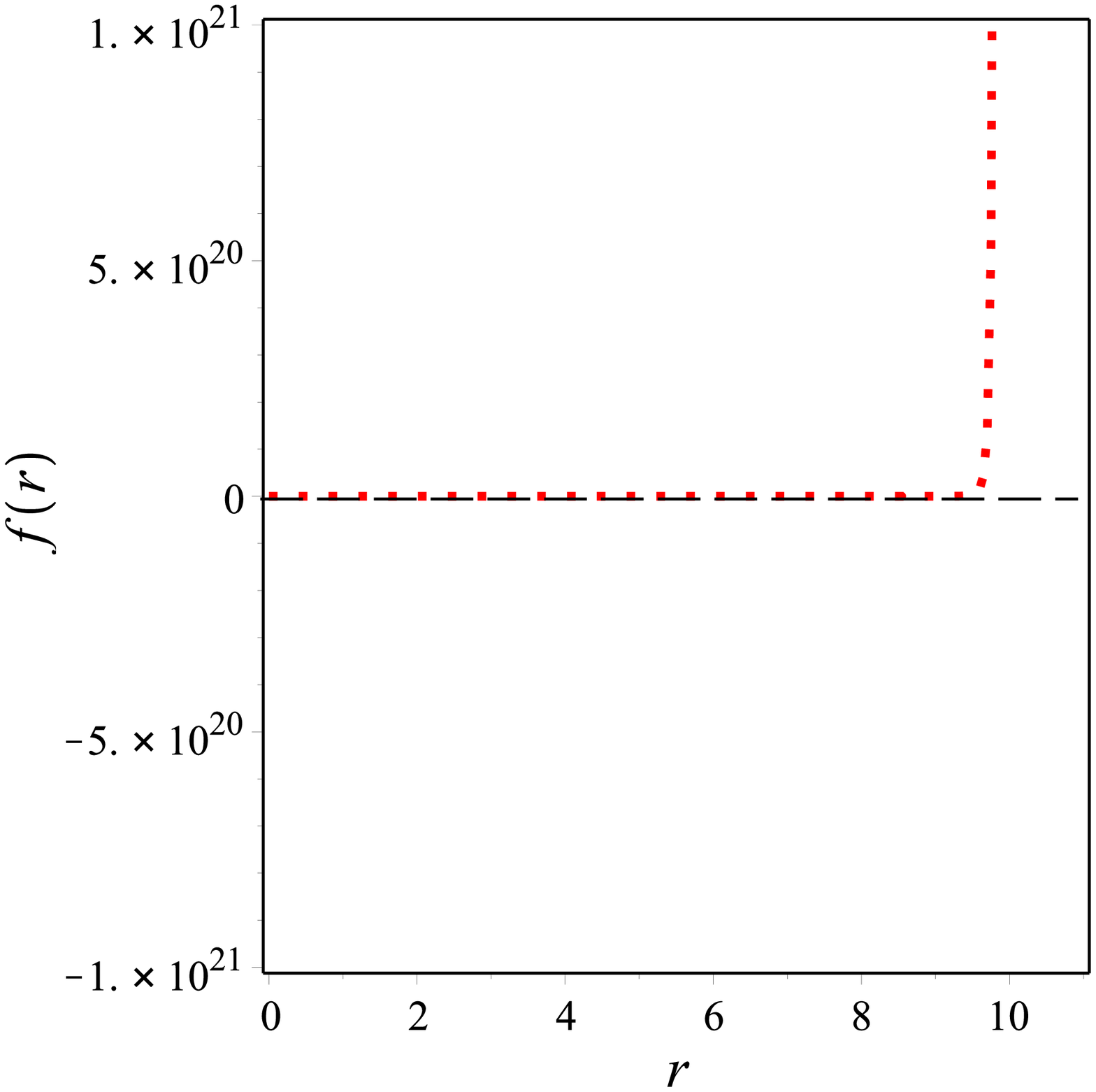}}
\subfigure[~The behavior of the function $f_R$  given   (\ref{fR2})]{\label{fig:fR}\includegraphics[scale=0.33]{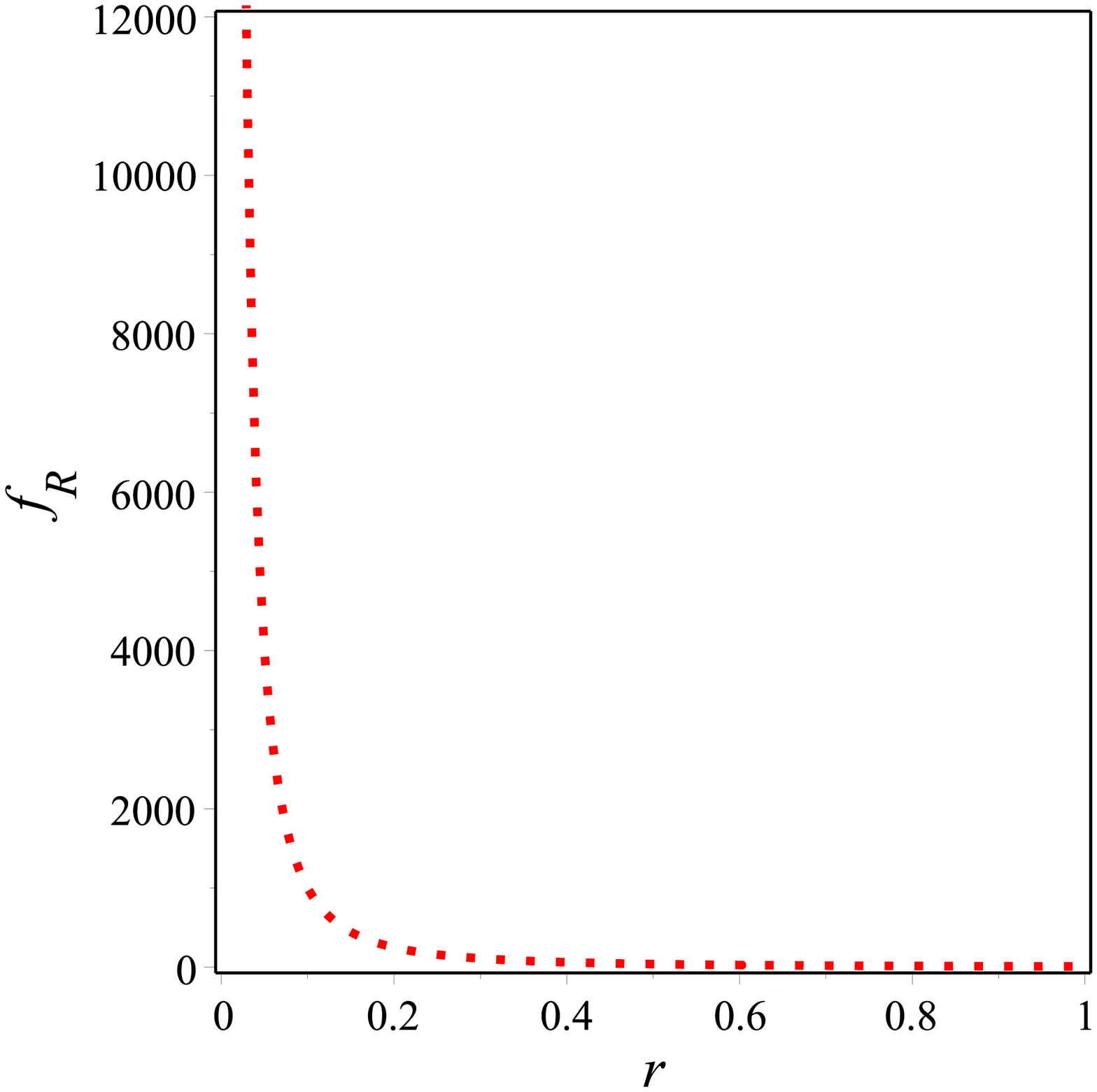}}
\subfigure[~The behavior of the function $f_{RR}$  given   (\ref{fR3})]{\label{fig:fRR}\includegraphics[scale=0.33]{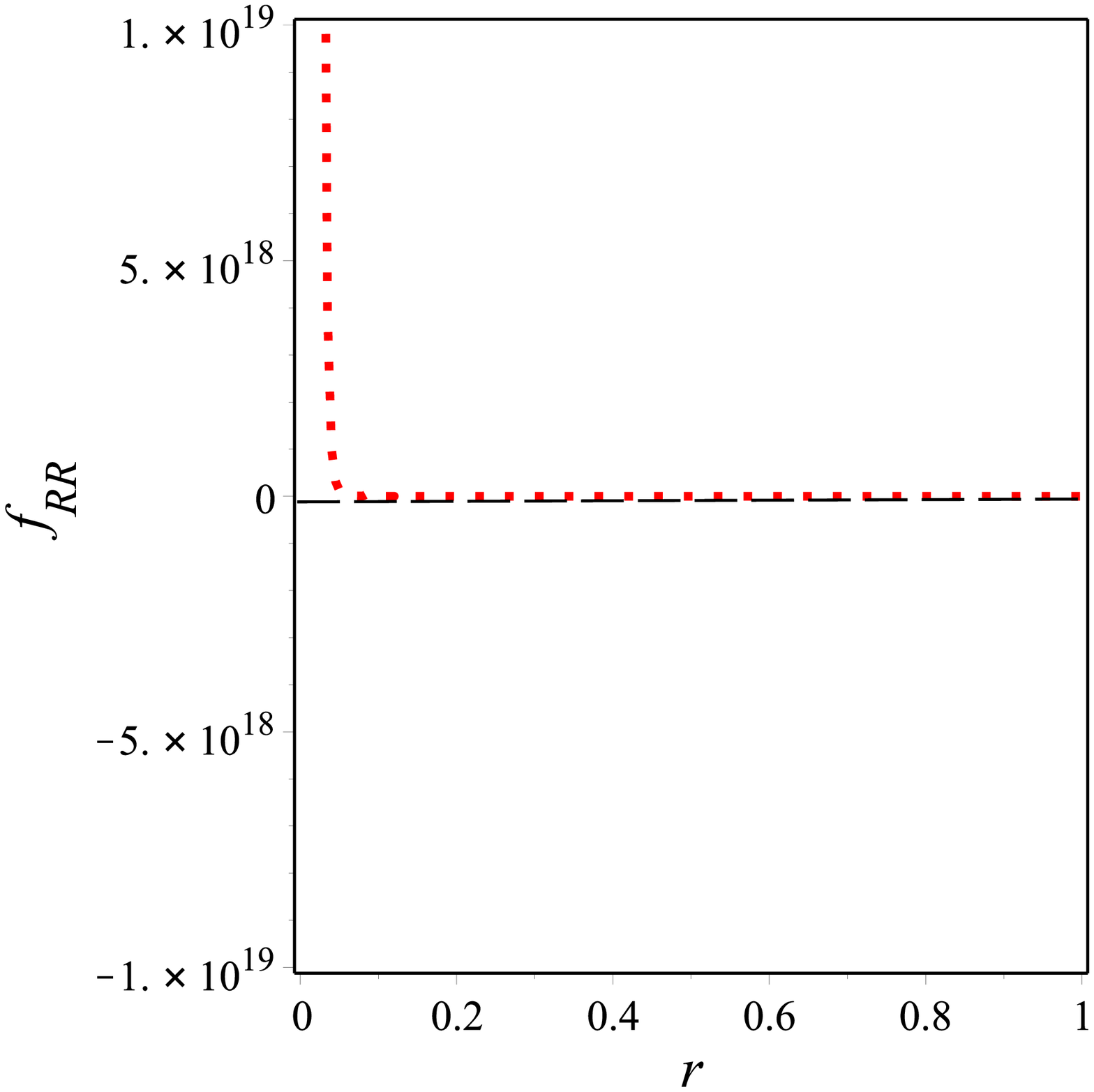}}
\caption[figtopcap]{\small{Systematic plots of; \subref{fig:R} the Ricci scalar given by Eq. (\ref{R1});   \subref{fig:fr} the analytic function f(r) given   (\ref{fR1});  \subref{fig:fR} the derivative function $f_R$ given by Eq. (\ref{fR2}), and  \subref{fig:fRR} the second derivative $f_{RR}$ given by Eq. (\ref{fR3}). All the figures  are plotted using the following values of the constants, $M=1$, $c_0=0.1$, $c_1=10$, $c_2=-10^{5}$. These values satisfy the constrains given by Eq. (\ref{cond}). }}
\label{Fig:1}
\end{figure}

The behavior of the Ricci scalar, $f(r)$, $f_R$ and $f_{RR}$ are given in Figure \ref{Fig:1}.
As figure~\ref{Fig:1}~\subref{fig:R}--\ref{Fig:1}~\subref{fig:fRR} shows that the Ricci scalar, $f(r)$, the first derivative of $f(R)$ and the second derivative of $f(R)$ all of them have positive value which means that the condition of stability given by Dolgov-Kawasaki  is satisfied \cite{DeFelice:2010aj,Bertolami:2009cd,Faraoni:2006sy,Cognola:2007zu}.
\newpage
\subsection{Physical properties of the rotating BH solution,   Eq. (\ref{sp3}) }
Now, let us turn our attention to the rotating case to where  the asymptote behaviors of the metric potentials, $g_{tt}(r)$, $g_{r\phi}(r)$, and $g_{rr}(r)$ are given by Eq. (\ref{sp3}) and get
 \begin{eqnarray}\label{mpabr}
&& g_{t\,t }=\frac{ \frac{24{r}^{6}{c_0}^{3}}{c_1} \left( 1-2c_4c_1 \right)\ln \Bigg(c_0+\frac{c_1}{r^2}\Bigg) +2{c_1}^{3}c_4- \left( 1+12 c_4c_0{r}^{2} \right) {c_1}^{2}+ 6c_0{r}^{2}\left(1+ 7c_4{c_0}{r}^{2} \right) c_1 + \left(  \left( {c_4}^{2} -c_2\right) {r}^{2}-21{c_0}^{2} \right) {r}^{4} }{{r}^{4}}\,,\nonumber\\
 && g_{t\,t }(r\rightarrow \infty)\approx r^2{\Lambda_{1}}_{eff}-{\mathcal M}+\frac{\mathcal J}{r^2}-\frac{\mathcal J_1{}^2}{r^4}-\mathcal{O}(r^{-6})\,, \nonumber\\
 &&  g_{t\,t }(r\rightarrow 0)\approx\Bigg({\Lambda_{1}}_{eff}-48c_0{}^3(1-2c_1c_4)\ln\,r\Bigg)r^2-7{\mathcal M}-\frac{24[2c_1c_4-1]c_0{}^4r^4}{c_1{}^2}+\mathcal{O}(r^{6})\,,\nonumber\\
 && g_{r\,r }=\frac{  g_{t\,t }}{\left(c_0-\frac{c_1}{r^2}\right)^6}\,,\nonumber\\
 && g_{r\,r }(r\rightarrow \infty)\approx \frac{c_1c_0{}^6}{(c_1c_2-24c_0{}^3\ln\,c_0[1-24c_0{}^3c_1\ln\,c_0])r^2}+\frac{3c_1{}^2c_0{}^5(2c_1c_2-c_0{}^3[1-48\ln\,c_0\{1+c_0{}^3c_1(1+24\ln\,c_0)\}])}
 {(c_1c_2-24c_0{}^3\ln\,c_0[1-24c_0{}^3c_1\ln\,c_0])^2r^4}\nonumber\\
 &&+\mathcal{O}(r^{-6})\,, \nonumber\\
 &&  g_{r\,r }(r\rightarrow 0)\approx\Bigg({\Lambda_{1}}_{eff}-48c_0{}^3(1-2c_1c_4)\ln\,r\Bigg)r^2-7{\mathcal M}-\frac{24[2c_1c_4-1]c_0{}^4r^4}{c_1{}^2}-\mathcal{O}(r^{6})\,,\nonumber\\
 && g_{t\,\phi }=\frac{c_2r^6-24c_0{}^3r^6\ln\left(c_0+\frac{c_1}{r^2}\right)+21c_0{}^2c_1r^4-6c_0c_1{}^2r^2+c_1{}^3}{r^4}\,,\nonumber\\
 &&
 g_{t\,\phi }(r \rightarrow \infty)\approx (c_4-24c_0{}^3\ln\,c_0)r^2-3c_0{}^2c_1+\frac{6c_0c_1{}^2}{r^2}-\frac{7c_1{}^3}{r^4}-\mathcal{O}(r^{-6})\nonumber\\
 &&g_{t\,\phi }(r \rightarrow 0)\approx 21c_0{}^2c_1+ \left[c_4+24c_0{}^3\ln\left(\frac{r^2}{c_1}\right)\right]r^2-\frac{24c_0{}^4 r^4}{c_1}+\mathcal{O}(r^{6})\,.\end{eqnarray}
where  ${\mathcal M}=-3c_0{}^2(1-2c_1c_4)$,\,\,\, ${\Lambda_{1}}_{eff}=\frac{24{c_0}^{3}}{c_1} \left( 1-2c_4c_1 \right)\ln\, c_0+c_4{}^2-c_2$, ${\mathcal J}=-6c_0c_1(1-2c_1c_4)$, $\mathcal J_1{}^2=-7c_1{}^2(1-2c_1c_4)$. Using Eq. (\ref{mpab}) in (\ref{met12}), when $b_2\neq0$, we get
\begin{eqnarray} \label{metafr}
& &  ds^2\approx -\Bigg[r^2{\Lambda_{1}}_{eff}-{\mathcal M}+\frac{{\mathcal J}}{r^2}-\frac{{\mathcal J_1{}^2}}{r^4}\Bigg]dt^2 +
\displaystyle\frac{dr^2}{\frac{c_1c_0{}^6}{(c_1c_2-24c_0{}^3\ln\,c_0[1-24c_0{}^3c_1\ln\,c_0])r^2}+\frac{3c_1{}^2c_0{}^5(2c_1c_2-c_0{}^3[1-48\ln\,c_0\{1+c_0{}^3c_1(1+24\ln\,c_0)\}])}
 {(c_1c_2-24c_0{}^3\ln\,c_0[1-24c_0{}^3c_1\ln\,c_0])^2r^4}}\nonumber\\
 &&+r^2d\phi^2+{\frac {c_4\,{r}^{6}+48 {r
}^{6}\,{c_0}^{3}\ln\,r-24\,{c_0}^{3}\ln  \left( c_0\,{r}^{2}+c_1
 \right) {r}^{6}+21\,{r}^{4}{c_0}^{2}c_1-6\,{r}^{2}c_0
\,{c_1}^{2}+{c_1}^{3}}{{r}^{4}}}\,dt\,d\phi\,.
\end{eqnarray}
The line element (\ref{metaf}) is asymptotically approaching AdS
spacetime and as we discussed in the non-rotating case it does not
coincide with the rotating BTZ spacetime \cite{Quevedo:2008ry} due
to the contribution of the extra terms that come from the
higher-order curvature terms of  $\mathit{f(R)}$
\cite{Setare:2003hm}. Now we are going to use Eq. (\ref{sp3})  in
Eq. (\ref{Ricci})  and get
\begin{eqnarray} \label{R1r}
&&R(r)=\frac {6}{c_1\, \left( c_0\,{r}^{2}+c_1 \right) ^{2 } \left( c_1-c_0\,{r}^{2} \right) ^{7}}\Bigg\{{c_0}^{3}c_2\,{r}^{18}c_1-3\,{c_0}^{2 }c_2\,{r}^{16}{c_1}^{2}-9\,c_0\,c_2\,{r}^{14}{ c_1}^{3}+48\,{c_0}^{6}\ln\,r {r}^{18}-1968\,{ r}^{12}{c_1}^{4}{c_0}^{6}-4\,{c_1}^{10}\nonumber\\
 &&+576\,{r}^{14}{c_1}^{3}{c_0}^{7}-172\,{r}^{12}{c_1}^{3}{c_0}^{3} -76\,{c_0}^{4}{r }^{14}{c_1}^{2}+268\,{c_0}^{4}{r}^{8}{c_1}^{6}+24\,{c_0}^{5}{r}^{16}c_1-2940\,{c_0}^{5}{r}^{10}{c_1}^{5 }+20\,c_0\,{r}^{2}{c_1}^{9} -5\,c_2\,{r}^{12}{c_1}^{4} \nonumber\\
 &&+24{r}^{8}\,\ln\,r[96\,{c_0 }^{8}{c_1}^{2}{r}^{8}-304\,{c_0}^{7}{c_1}^{3}{r}^{6}-688{c_0}^{6}{c_1}^{4}{r}^{4}-144\,{c_0}^{5}{c_1}^{5}{r}^ {2}-18\,{ c_0}^{4}{c_1}^{2}{r}^{6} -10\,{c_0}^{3}{c_1}^{3}{r}^{4} -6\,{c_0}^{ 5}c_1\,{r}^{8}+16\,{c_0}^{4}{c_1}^{6}]\nonumber\\
 &&+24{c_0}^{4}r^8c_1\ln  \left( c_0\,{r}^{2}+c_1 \right)\left[9\,{r}^{6}{c_1}+152{r}^{6}{c_1}^2{c_0}^{3}+344 {r}^ {4}{c_1}^3{c_0}^{2}+72{c_0}{r}^{2}{c_1}^4-192{c_1}^5+3 c_0{r}^{8}+5{r}^{4}{c_1}^2- {c_0} ^{2}{r}^{10}\right]\nonumber\\
 &&-576c_0{}^6c_1r^{12}\ln \left(\left( c_0\,{r}^{2}+c_1 \right)\right)^2\left[3 {r}^{4}{c_0}^{2}c_1+9{r }^2{c_1}{}^2{c_0}+5{c_1}^{3}{r}^2{c_0}^3 \right]-24\,{c_0}^{3}{r}^{6}{c_1}^{7}-24 \,{c_0}^{2}{r}^{4}{c_1}^{8}-36\,{c_0}^{2}{r}^{10}{c_1}^{4}+4\,c_0\,{r}^{8}{c_1}^{5}\nonumber\\
 &&- 2304{c_0}^{6}c_1\left( \ln\,r \right) ^{2}r^{12} \left[3\,{c_0}^{2}{c_1}{r}^{4}+9\,{c_0}{c_1}^{2}{r}^{2}+5\,{c_1}^{3}-{r}^{6}{ c_0}^{3}\right]+1152r^8c_0{}^7c_1\ln \left( c_0\,{r}^{2}+c_1 \right)\ln\,r \Bigg[18\,{c_1}^{2}{r}^{6}+ 6{c_0}{c_1}{r}^{8} -2{r}^{10}{c_0}^{2}\nonumber\\
 &&- {r}^{6}{c_0}^{3}+10{c_1}^{3}{r}^{4}\Bigg]\Bigg\}
\,,\nonumber\\
 &&R(r\rightarrow \infty) \approx -\frac {6[c_2c_1+576\,{c_0}^{6} \left( \ln  c_0\right) ^{2}-24\,{c_0}^{3}c_3\,\ln  c_0] }{{c_0}^{6}c_1}-\frac { 12\left( c_2c_1+576\,{c_0}^{6}c_1 \left( \ln
 c_0  \right) ^{2}-24\,{c_0}^{3}\,
\ln  c_0  \right)}{{c_0}^{7}{r}^{2}}\nonumber\\
 &&+\frac{6{c_1} \left( 7\,c_2+4032\,{c_0}^{6}
 \left( \ln c_0 \right) ^{2}-168\,{c_0}^{3}
c_3\,\ln  c_0 -8\,{c_0}^{3}c_3+
384\,{c_0}^{6}\ln  c_0 \right) }{{c_0}^{8}{r}^{4}}
+\mathcal{O}(r^{-6})\,,\nonumber\\
 &&R(r\rightarrow 0) \approx -24+\frac{48c_0{}^2r^{4}}{c_1{}^2}+\mathcal{O}(r^{6})\,,\nonumber\\
 &&r(R)\approx \pm2\sqrt{\frac {3c_1(c_2c_1+576\,{c_0}^{6}c_1 \left( \ln
 c_0  \right) ^{2}-24\,{c_0}^{3}\,
\ln  c_0)}{c_0(144\,{c_0}^{3}c_4 \ln c_0-R\,{c_0}^{6}c_1-6\,c_2c_1-3456\,{c_0}^{6}c_1 \left(
\ln c_0  \right) ^{2})}}\,,
\qquad \qquad  \qquad \qquad r\rightarrow \infty\,.
\end{eqnarray}
Equation (\ref{R1}) shows that when the constant $c_0$  must not equal to zero also Eq. (\ref{R1}) shows that either
\begin{eqnarray} \label{cond3}
&& c_0>0\,, \qquad {\textrm and} \qquad c_1>0\,, \qquad {\textrm and} \qquad (c_2c_1+576\,{c_0}^{6}c_1 \left( \ln
 c_0  \right) ^{2}-24\,{c_0}^{3}\,
\ln  c_0)>0\,, \nonumber\\
 &&{\textrm and} \qquad (144\,{c_0}^{3}c_4 \ln c_0-R\,{c_0}^{6}c_1-6\,c_2c_1-3456\,{c_0}^{6}c_1 \left(
\ln c_0  \right) ^{2})>0\,,\nonumber\\
&&{\textrm or} \qquad c_1<0\,, \qquad {\textrm and} \qquad (c_2c_1+576\,{c_0}^{6}c_1 \left( \ln
 c_0  \right) ^{2}-24\,{c_0}^{3}\,
\ln  c_0)<0\,, \qquad {\textrm and}\nonumber\\
&& (144\,{c_0}^{3}c_4 \ln c_0-R\,{c_0}^{6}c_1-6\,c_2c_1-3456\,{c_0}^{6}c_1 \left(
\ln c_0  \right) ^{2})>0\,,\nonumber\\
&&{\textrm or} \qquad c_1<0\,, \qquad {\textrm and} \qquad (c_2c_1+576\,{c_0}^{6}c_1 \left( \ln
 c_0  \right) ^{2}-24\,{c_0}^{3}\,
\ln  c_0)>0\,, \qquad {\textrm and}\nonumber\\
&& (144\,{c_0}^{3}c_4 \ln c_0-R\,{c_0}^{6}c_1-6\,c_2c_1-3456\,{c_0}^{6}c_1 \left(
\ln c_0  \right) ^{2})<0\nonumber\\
&&{\textrm or} \qquad c_1>0\,, \qquad {\textrm and} \qquad (c_2c_1+576\,{c_0}^{6}c_1 \left( \ln
 c_0  \right) ^{2}-24\,{c_0}^{3}\,
\ln  c_0)<0\,, \qquad {\textrm and}\nonumber\\
&& (144\,{c_0}^{3}c_4 \ln
c_0-R\,{c_0}^{6}c_1-6\,c_2c_1-3456\,{c_0}^{6}c_1 \left( \ln c_0
\right) ^{2})<0\,.\end{eqnarray}

The  form of  $f(r)$ of the BH solution (\ref{sp3}) has the
following form:
\begin{eqnarray} \label{fR1r}
&&f(r)=\frac {1}{ \left( c_0\,{r}^{2}+c_1 \right) {c_0}^{5} \left( -c_0\,{r}^{2}+c_1 \right) ^{7}c_1}\Bigg\{4\,{c_1}^{9}c_2-96\,{c_0}^{11}{r}^{16}\ln  \left( c_0\,{r}^{2}+c_1\right) +c_5\,{c_0}^{5}{c_1}^{9}+12\,{c_0}^{7}{r}^{14}c_2\,{c_1}^{2}- 24\,{c_1}^{8}{r}^{2}c_0\,c_2\nonumber\\
 && +56\,{c_1}^{7}{r}^{4 }{c_0}^{2}c_2-56\,{c_1}^{6}{r}^{6}{c_0}^{3}c_2+20\,{c_1}^{4}{c_0}^{5}{r}^{10}c_2-108\,{c_1}^{3}{c_0}^{6}{r}^{12}c_2+192\,{c_0}^{11}{r}^{16}\ln\,r -14\,c_5\,{c_0}^{8}{c_1}^{6 }{r}^{6}+96\,{c_0}^{10}{r}^{14}c_1 \nonumber\\
 && +864\,{c_1}^{3}{c_0}^{8}{r}^{10}\ln  \left( c_0 \,{r}^{2}+c_1 \right)+1248\,{c_1}^{2}{c_0}^{9}{r}^{12} \ln  \left( c_0\,{r}^{2}+c_1 \right)+288\,{c_0}^{10}{r }^{14}c_1\,\ln  \left( c_0\,{r}^{2}+c_1 \right) -4608\, \ln  \left( c_0\,{r}^{2}+c_1 \right) {c_0}^{13}{c_1}^{2}{r}^{14}\nonumber\\
 && -20736\,{r}^{10}{c_0}^{11} \left( \ln  \left( c_0\,{r}^{2}+c_1 \right)  \right) ^{2}{c_1}^{4}-29952\,{r}^{ 12}{c_0}^{12} \left( \ln  \left( c_0\,{r}^{2}+c_1 \right)  \right) ^{2}{c_1}^{3}-6912\,{r}^{14}{c_0}^{13} \left( \ln  \left( c_0\,{r}^{2}+c_1 \right)  \right) ^{2}{c_1}^{2}\nonumber\\
 &&+2304\,{r}^{16}{c_0}^{14} \left( \ln  \left( c_0 \,{r}^{2}+c_1 \right)  \right) ^{2}c_1+13824\,\ln  \left( c_0\,{r}^{2}+c_1 \right) {c_0}^{10}{c_1}^{5}{r}^{8} -1536\,\ln  \left( c_0\,{r}^{2}+c_1 \right) {c_0}^{9}{c_1}^{6}{r}^{6}+480\,{c_1}^{8}{r}^{2}{c_0}^{7}\nonumber\\
 &&+52224\,\ln  \left( c_0\,{r}^{2}+c_1 \right) {c_0}^{11}{c_1}^{4}{r}^{10}+13824\,\ln  \left( c_0\,{r}^{2}+c_1 \right) {c_0}^{12}{c_1}^{3}{r}^{12 }-1280\,{c_1}^{7}{r}^{4} {c_0}^{8}+2880\,{c_1}^{6}{r}^{6}{c_0}^{9}-21696\,{c_1}^{5}{c_0}^{10}{r}^{8}\nonumber\\
 &&-119808\,{r}^{12}{c_0}^{12} \left( \ln \,r \right) ^{2}{c_1}^{3}-27648\,{r}^{14}{c_0}^{13} \left( \ln  \,r  \right) ^{2}{c_1}^{2} +9216\,{r}^{16}{c_0}^{14} \left( \ln \,r  \right) ^ {2}c_1+3072\,{c_1}^{6}{c_0}^{9}{r }^{6}\ln  \,r-27648\,{c_1}^{5}{c_0}^{10}{r}^{8 }\ln \,r\nonumber\\
 &&  -104448\,{c_1}^{4}{c_0}^{11}{r}^{10}\ln \,r - 27648\,{c_1}^{3}{c_0}^{12}{r}^{12}\ln \,r+ 9216\,{c_1}^{2}{c_0}^{13}{r}^{14}\ln \,r  -288 \,{c_1}^{4}{c_0}^{7}{r}^{8}-7200\,{c_1}^{4}{c_0}^{ 11}{r}^{10}+2304\,{c_1}^{3}{c_0}^{12}{r}^{12}\nonumber\\
  &&+\ln  \left( c_0\,{r}^{2}+c_1 \right)\left[27648\, {c_0 }^{13}{c_1}^{2}\ln\,r {r}^{14}-9216\, {c_0}^{ 14}c_1\,{r}^{16}\ln\,r +119808\,\ln {c_0}^{12}{c_1}^{ 3}{r}^{12}\ln \,r+82944\,\ln  {c_0}^{11}{c_1}^{4}{r }^{10}\ln  \,r \right]\nonumber\\
 &&-64\,{c_1}^{9}{c_0}^{6}  +14\,c_5\,{c_0}^{7}{c_1}^{7}{r}^{4}-6\,c_5\,{c_0}^{6}{c_1}^{8}{r}^{2}+14\,c_5\,{c_0}^ {10}{c_1}^{4}{r}^{10}-14\,c_5\,{c_0}^{11}{c_1}^{ 3}{r}^{12}+6\,c_5\,{c_0}^{12}{c_1}^{2}{r}^{14}-82944\,{r}^{10 }{c_0}^{11} \left( \ln  \,r  \right) ^{2}{c_1}^ {4}\nonumber\\
 &&-c_5\,{c_0}^{13}c_1\,{r}^{16}-1728\,{c_0}^{8}\ln \,r {c_1}^{3}{r}^{10}-2496\,{c_0}^{9}\ln\,r {c_1}^{2}{r}^{12}-576\,{c_0}^{10}c_1\,{r}^{14}\ln\,r+32\,{c_1}^{5}{r}^{6}{c_0}^{6}-1088\,{c_1}^{3}{c_0}^{8}{r}^{10}-288\,{c_1}^{2}{ c_0}^{9}{r}^{12} \nonumber\\
 && +96\ln  \left(\frac{2}{3}\right)c_0{}^3\left[14\,{c_1}^{5}{r}^{6}{c_0}^{3}-{c_1}^{8}-6\,{c_0}^{7} {r}^{14}c_1 -14\,{c_0}^{2}{c_1}^{6}{r}^{4} +6\,{c_0}{c_1}^{7}{r}^{2} -14\,{c_1}^{3}{c_0}^{5}{r}^{10} +{c_0}^{8}{r}^{16} +14\,{c_1}^{2}{c_0}^{6}{r}^{12}\right]\Bigg\}
\,,\nonumber\\
 && f(r\rightarrow \infty)\approx\frac {1}{{c_0}^{7}c_1\,{r}^{4}}\Bigg\{96\,{c_0}^{5}{r}^{4}\ln  \left( \frac{3}{2} \right) -96\,{c_0}^{5}{r}^{4}\ln  \,c_0 +2304\,{c_0}^{8}{r}^{4}c_1\, \left( \ln \,c_0 \right) ^{2}-{c_0}^{7}{r}^{4}c_5\,c_1 -288\,c_1\,{r}^{2}{c_0}^{4}\ln\,c_0 \nonumber\\
 &&-c_5\,{c_0}^{13}c_1\,{r}^{16}-1728\,{c_0}^{8}\ln \,r {c_1}^{3}{r}^{10}-2496\,{c_0}^{9}\ln\,r {c_1}^{2}{r}^{12}-576\,{c_0}^{10}c_1\,{r}^{14}\ln\,r+32\,{c_1}^{5}{r}^{6}{c_0}^{6}-1088\,{c_1}^{3}{c_0}^{8}{r}^{10}-288\,{c_1}^{2}{ c_0}^{9}{r}^{12} \nonumber\\
 &&+6912\,{c_1}^{2}{r}^{2}{c_0}^{7} \left( \ln \,c_0 \right) ^{2}+12\,{c_1}^{2}{r}^{2}c_0\,c_2+864\,{c_1}^{2}{c_0}^{3}\ln\,c_0 -20736\,{c_1}^{3}{c_0}^{6} \left( \ln\,c_0 \right) ^{2}-36\,{c_1}^{3}c_2-2304\,{c_1}^{3}{c_0}^{6}\ln\,c_0 +48\,{c_0}^{3}{c_1}^{2}\Bigg\}\nonumber\\
 &&-c_5\,{c_0}^{13}c_1\,{r}^{16}-1728\,{c_0}^{8}\ln \,r {c_1}^{3}{r}^{10}-2496\,{c_0}^{9}\ln\,r {c_1}^{2}{r}^{12}-576\,{c_0}^{10}c_1\,{r}^{14}\ln\,r+32\,{c_1}^{5}{r}^{6}{c_0}^{6}-1088\,{c_1}^{3}{c_0}^{8}{r}^{10}-288\,{c_1}^{2}{ c_0}^{9}{r}^{12} \nonumber\\
 &&+\mathcal{O}(r^{-6})\,,\nonumber\\
 && f(r\rightarrow0)\approx \frac {4\,c_2\,{c_1}^{2}+96\,c_1\,{c_0}^{3}\ln
 \left( \frac{3}{2} \right) +c_5\,{c_0}^{5}{c_1}^{2}-64\,{c_0}^{6}{c_1}^{2}+96\,c_1\,{c_0}^{7}{r}^{2}+192\,{r}^{4}{c_0}^{
8}}{{c_0}^{5}{c_1}^{2}} +\mathcal{O}(r^{6})\,.
\end{eqnarray}
To avoid the tachyonic instability, we check the Dolgov-Kawasaki
stability criterion
\cite{DeFelice:2010aj,Bertolami:2009cd,Faraoni:2006sy,Cognola:2007zu}
which states that the second derivative of the gravitational model
$f_{RR}$ must be always positive. Using the chain rule we get
\begin{eqnarray} \label{fR2r}
&&f_R=\frac{df(R)}{dR}=\frac{df(r)}{dr}\frac{dr}{dR}=c_0+\frac{c_1}{r^2}\,
\end{eqnarray}
\begin{eqnarray} \label{fR3r}
&&f_{RR}=\left( c_0\,{r}^{2}+c_1 \right) ^{3} \left(c_0\,{r}^{2} - c_1 \right) ^{8}c_1\Bigg\{12{r}^{6} \Bigg[ \ln  \left( c_0\,{r}^{2}+c_1 \right)\left[ 24\, {c_0}^{7}{r}^{16}- 101376 {c_0}^{7}{ c_1}^{4}{r}^{10}\ln\,r-34560{c_0}^{6}{c_1}^{5}{r}^{8}\ln \,r\right] -8\,{c_0}^{2}{c_1 }^{9}\nonumber\\
 &&-{c_0}^{4}c_2\,{r}^{16}c_1-888\,{r}^{12}{c_1}^{3}{c_0}^{8}+6480\,{r}^{10}{c_1}^{4}{c_0}^{7}+15848 \,{r}^{8}{c_1}^{5}{c_0}^{6}+656\,{r}^{8}{c_1}^{4}{c_0}^{3}-32\,{c_0}^{6}{r}^{14}c_1+7712\,{c_0}^{5}{c_1}^{6}{r}^{6}+280\,{c_0}^{5}{r}^{12}{c_1}^{2}\nonumber\\
 &&-488\,{c_0}^{4}{c_1}^{7}{r}^{4}+816\,{c_0}^{4}{r}^{10}{c_1}^{3 }+12\,{c_0}^{3}c_2\,{r}^{14}{c_1}^{2}+42\,{c_0}^ {2}c_2\,{r}^{12}{c_1}^{3}+44\,c_0\,c_2\,{r}^{ 10}{c_1}^{4}+15\,c_2\,{r}^{8}{c_1}^{5}+16\,{c_0} ^{3}{r}^{2}{c_1}^{8}+80\,{c_0}^{2}{c_1}^{5}{r}^{6}\nonumber\\
 && +2304r^8c_0{}^6c_1\left( \ln\,r\right) ^{2}\Bigg[44 \,{c_0}{c_1}^{3}{r}^{2} +15{c_1}^{4}-{r}^{8}{c_0}^{4}+12{c_0}^{3}{c_1}{r}^{6} +42{c_0}^{2}{c_1}^{2}{r}^{4} \Bigg]+48r^4c_0{}^3c_1\ln \,r\Bigg[44\,{c_0}{c_1}^{2}{r}^{6} -16\,{c_0}{c_1}^{6}+15{c_1}^{3}{r}^{4}\nonumber\\
 &&-64\,{c_0}^{6}{c_1}{r}^{10}+560\,{c_0}^{5}{c_1}^{2}{r}^{8}+1632\,{c_0}^{4}{c_1}^{3}{r}^{6}+12\,{c_0}^{3}{r}^{10} +160\,{c_0}^{2}{c_1}^{5}{r}^{2} +42\,{c_0}^{2}{c_1}{r}^{8}+1312\,{c_0}^{3}{c_1}^{4}{r}^{4}\Bigg]-8c_0\,{c_1}^{6}{r}^{4}\nonumber\\
 &&   +24r^4c_1c_0{}^3\ln  \left( c_0\,{r}^{2}+c_1 \right)\Bigg[64 {r}^{10}{c_1}{c_0}^{6}-560{r}^{8}{c_1}^{2}{c_0}^{5}- 1632{r}^{6}{c_1}^{3}{c_0}^{4}-1312{r}^{4}{c_1}^{4}{c_0}^{3}-12 {c_0}^{3}{r}^{8}r^2-160\ {c_0}^{2}{c_1}^{5 }-42 {c_0}^{2}{r}^{8}{c_1}\nonumber\\
 &&+16 {c_0}{c_1}^{6}-44 {c_0}{r}^{6}{c_1}^{2} -15{r}^{4}{c_1} ^{3}-1152 {c_0}^{6}{c_1}{r}^{10}\ln \,r \Bigg]+576c_1c_0{}^6r^8 \left( \ln  \left( c_0\,{r}^{2}+c_1 \right)  \right) ^{2}\Bigg[12\,{r}^{6}{c_1}{c_0}^{3}+42\, {r}^{4}{c_1}^{2}{c_0}^{2}+44\, {r}^{2}{c_1}^{3}{c_0}\nonumber\\
 &&+15{c_1}^{4}-{r}^{8}{c_0}^{4}\Bigg]-48 \,{c_0}^{7}{r}^{16}\ln \,r-96768\,\ln  \left(c_0\,{r}^{2}+c_1 \right) {c_0}^{8}{c_1}^{3}{r}^{12} \ln \,r +2304\,\ln  \left( c_0\,{r}^{2}+c_1 \right) {r}^{16}\ln \,r c_1\,{c_0}^{10} \Bigg] \Bigg\}^{-1}
\,.
\end{eqnarray}

The behavior of the Ricci scalar, $f(r)$, $f_R$ and $f_{RR}$ are given in Figure \ref{Fig:2}.
Following the same procedure of the non-rotating we get the  behavior of the Ricci scalar, $f(r)$, $f_R$ and $f_{RR}$, in the rotating case,  as:
\begin{figure}
\centering
\subfigure[~The behavior of the Ricci scalar given by Eq. (\ref{R1r})]{\label{fig:Rr}\includegraphics[scale=0.33]{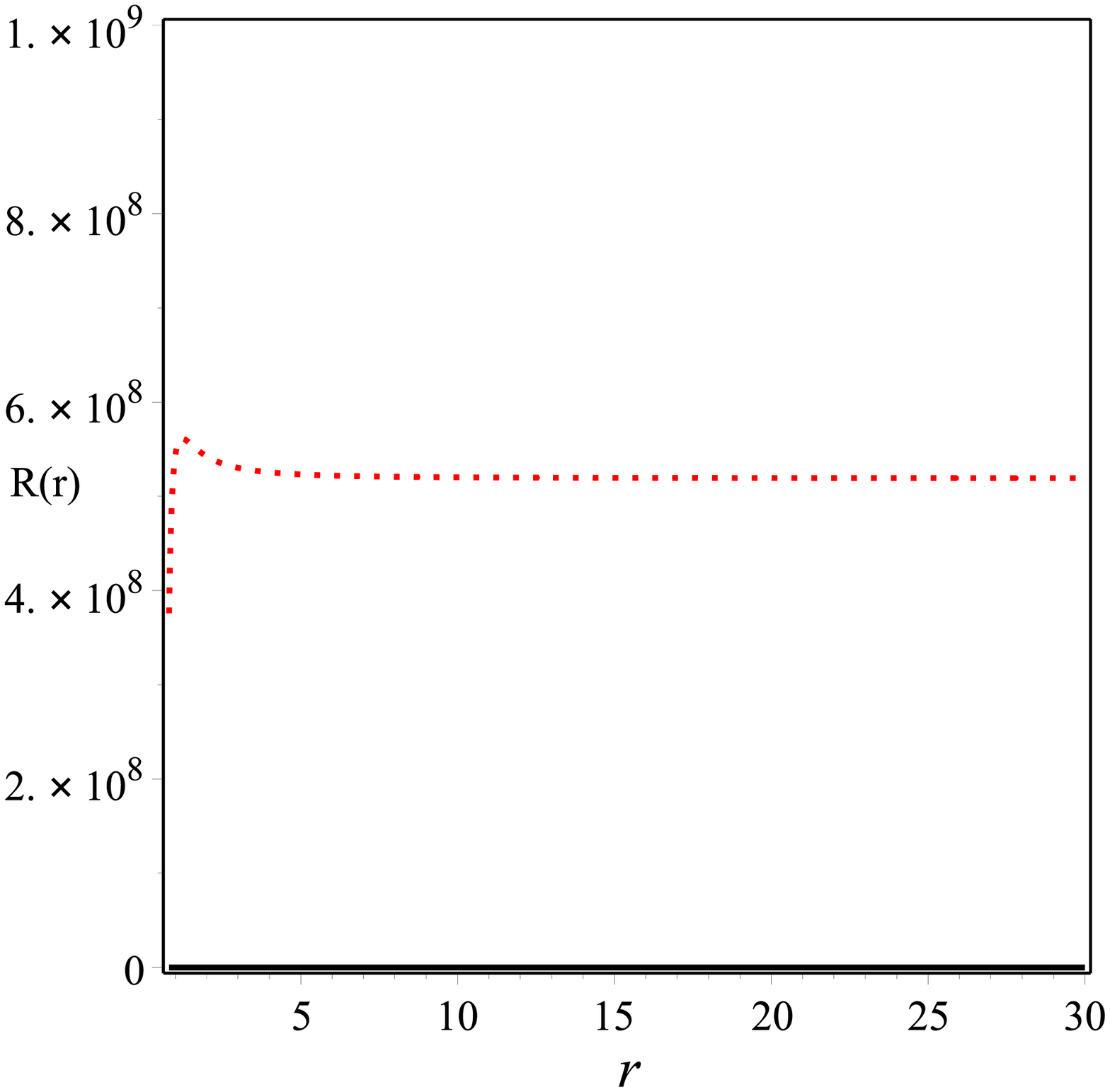}}
\subfigure[~The behavior of the function $f(r)$  given   (\ref{fR1r})]{\label{fig:frr}\includegraphics[scale=0.33]{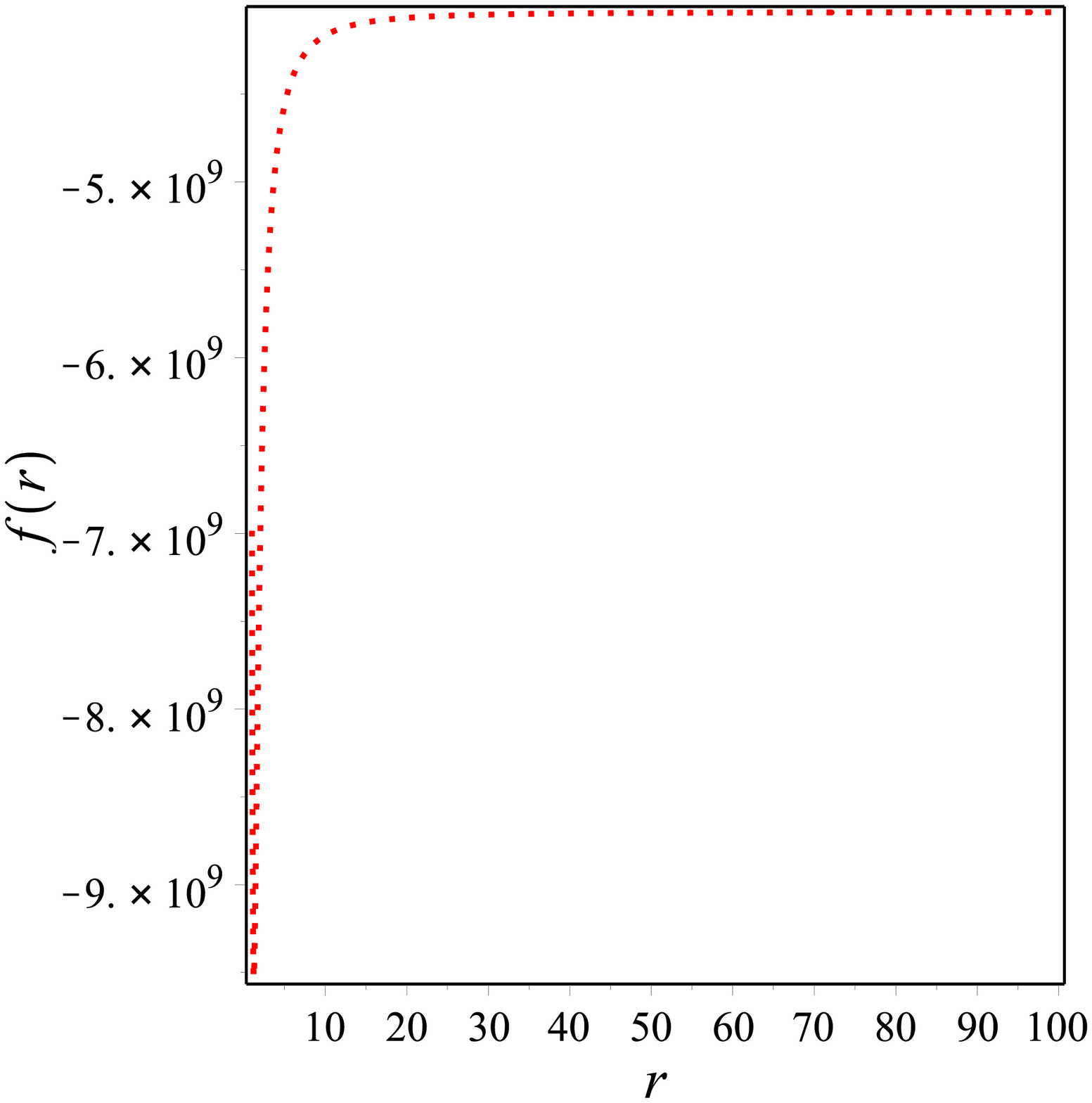}}
\subfigure[~The behavior of the function $f_R$  given   (\ref{fR2r})]{\label{fig:fRr}\includegraphics[scale=0.33]{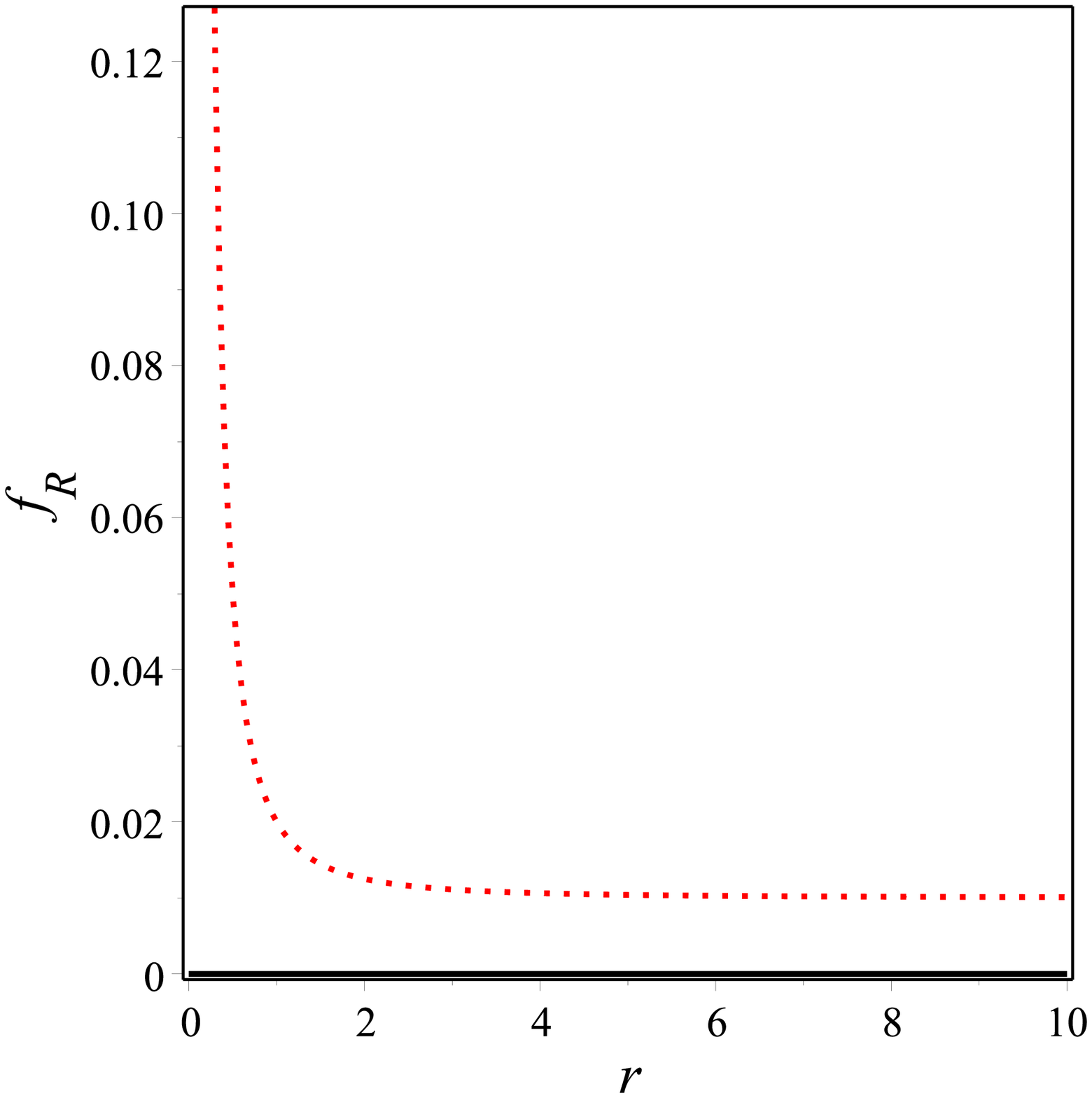}}
\subfigure[~The behavior of the function $f_{RR}$  given   (\ref{fR3r})]{\label{fig:fRRr}\includegraphics[scale=0.33]{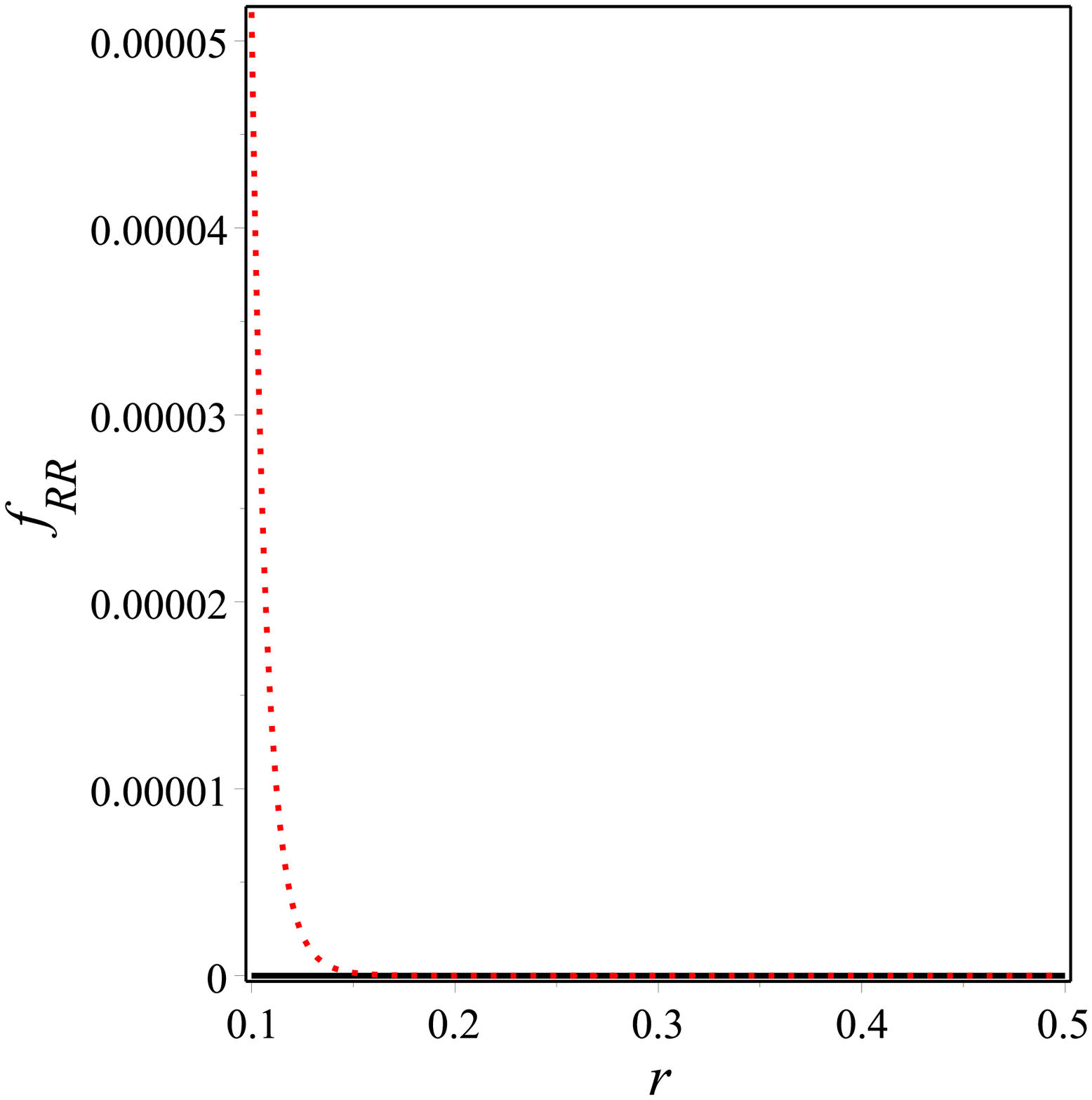}}
\caption[figtopcap]{\small{Systematic plots of; \subref{fig:Rr} the Ricci scalar given by Eq. (\ref{R1r});   \subref{fig:frr} the analytic function f(r) given   (\ref{fR1r});  \subref{fig:fRr} the derivative function $f_R$ given by Eq. (\ref{fR2r}), and  \subref{fig:fRRr} the second derivative $f_{RR}$ given by Eq. (\ref{fR3r}). All the figures  are plotted using the following values of the constants, $c_0=10^2$, $c_1=10$, $c_2=-10^{5}$, $c_3=1$ and $c_4=1$. These values satisfy the constrains given by Eq. (\ref{cond}). }}
\label{Fig:2}
\end{figure}
As
figure~\ref{Fig:2}~\subref{fig:Rr}--\ref{Fig:2}~\subref{fig:fRRr}
shows that the Ricci scalar has a positive value, $f(r)$ has a
costive value then a non-defined value then a negative value; the
first derivative of $f(R)$ has a positive value as well as the
second derivative of $f(R)$.
\section{Thermodynamics of the BHs}\label{S5}
Now, we are ready to study the physical properties the BHs
(\ref{sp1}) and (\ref{sp3}) from the thermodynamics viewpoint. To
do this, we are going to write the basic definitions of the
thermodynamical quantities  that we will use.
\subsection{Thermodynamics of the BH  (\ref{sp1})}\label{S41}
\begin{figure}
\centering
\subfigure[~The behavior of the metric potential given by Eq. (\ref{sp1})]{\label{fig:metpot}\includegraphics[scale=0.23]{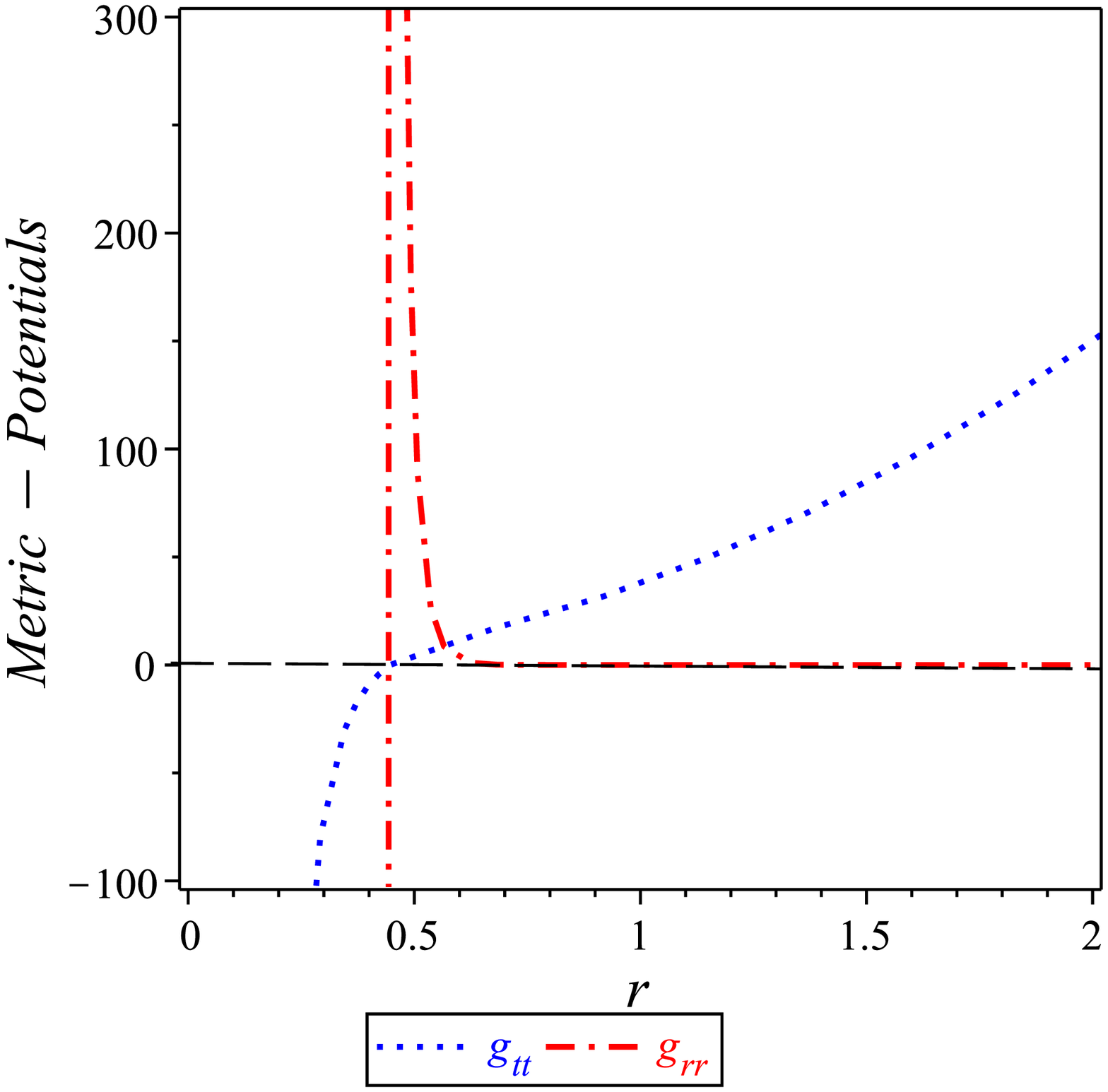}}
\subfigure[~The behavior of the horizons  given   by Eq.  (\ref{hor1})]{\label{fig:hor}\includegraphics[scale=0.23]{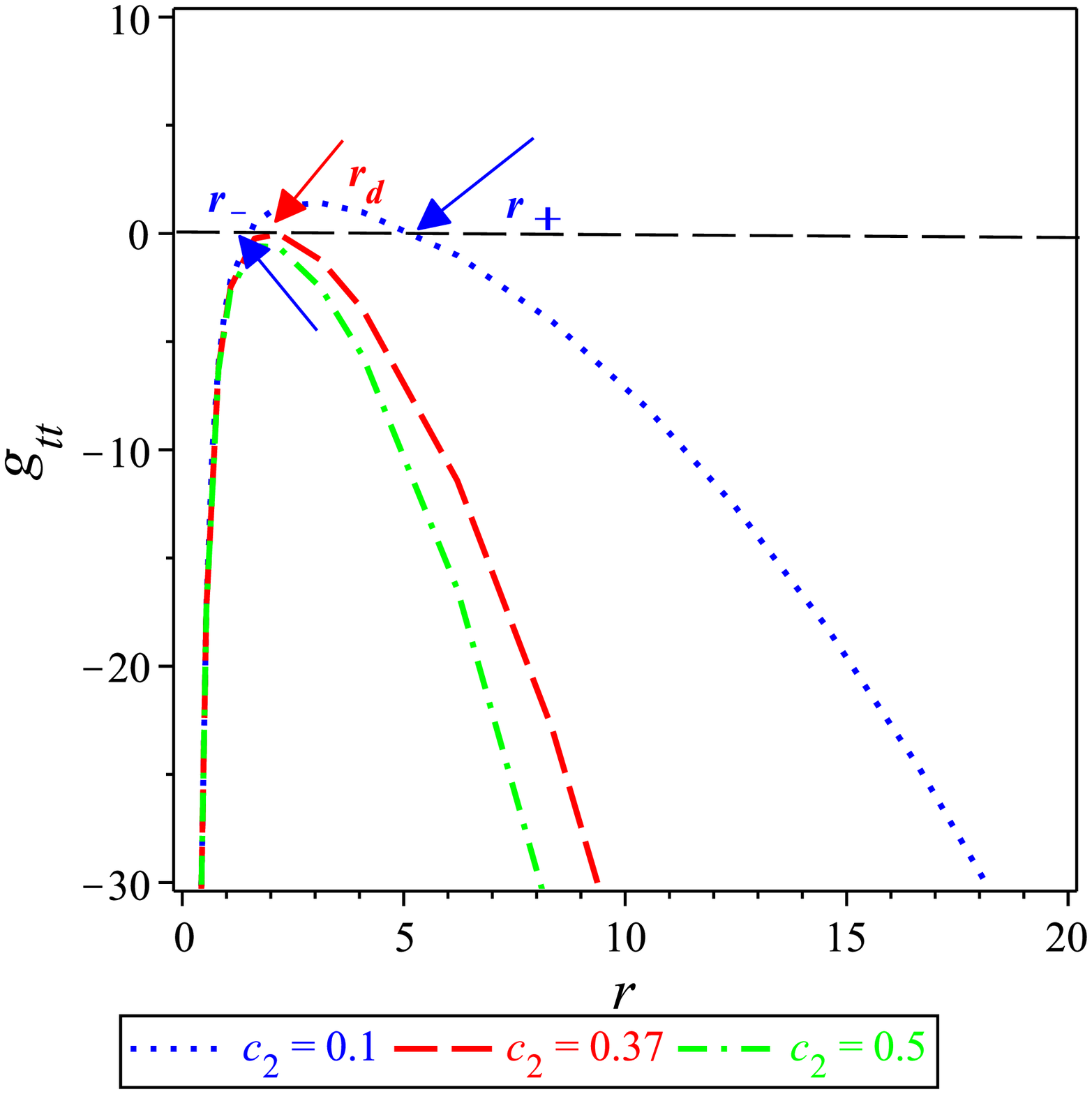}}
\subfigure[~The behavior of the Hawking temperature  given   (\ref{temp})]{\label{fig:temp}\includegraphics[scale=0.23]{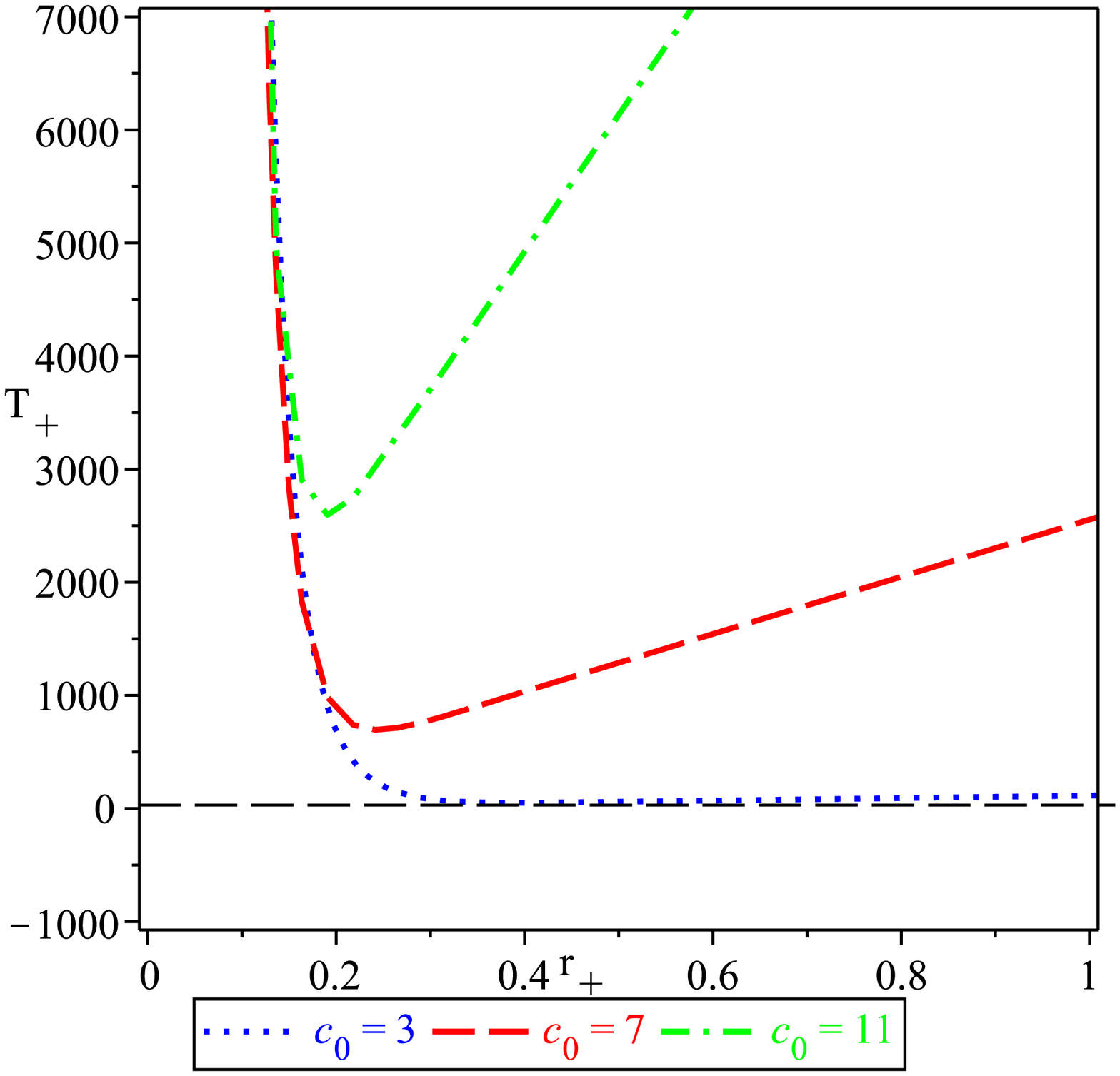}}
\subfigure[~The behavior of  Bekenstein-Hawking entropy  given   (\ref{ent})]{\label{fig:ent1}\includegraphics[scale=0.23]{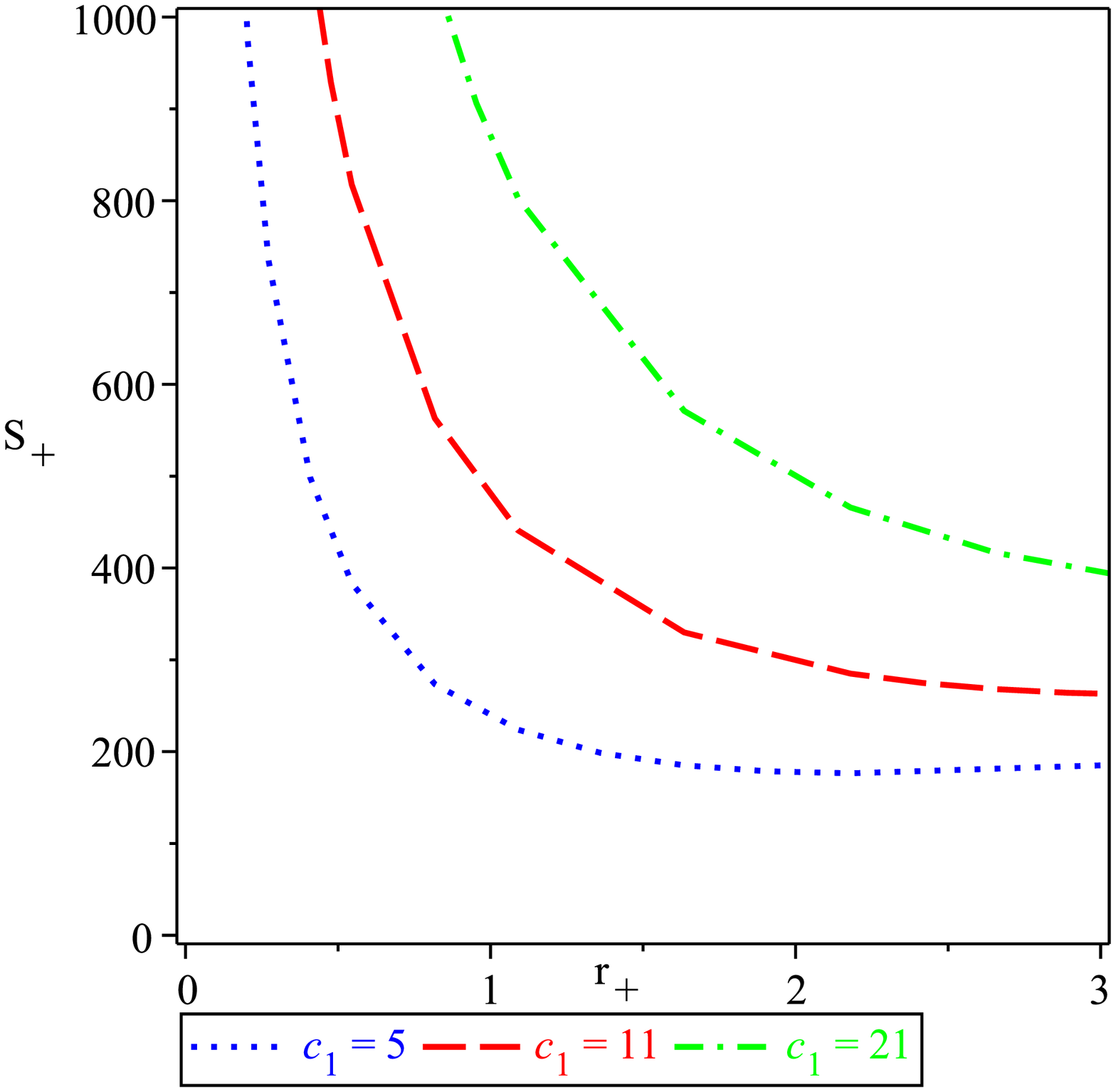}}
\subfigure[~The behavior of the heat capacity  given   (\ref{heat})]{\label{fig:heat1}\includegraphics[scale=0.23]{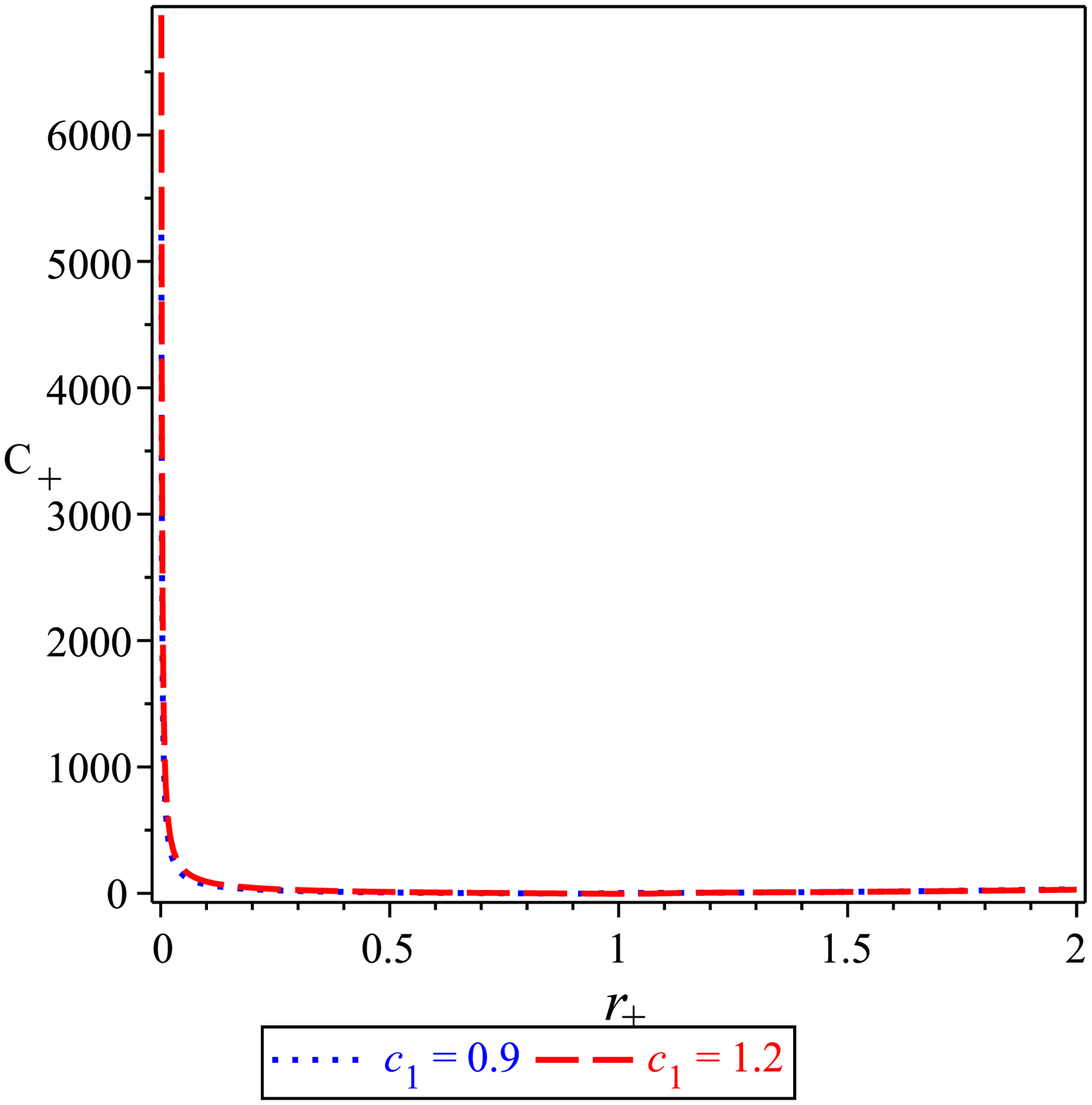}}
\caption[figtopcap]{\small{Systematic plots of; \subref{fig:metpot} the metric potentials  given by Eq. (\ref{sp1});   \subref{fig:hor} the horizons of the temporal component of $g_{tt}(r)$ given   (\ref{hor});  \subref{fig:temp} the Hawking temperature given by Eq. (\ref{temp}), and  \subref{fig:fRRr} the second derivative $f_{RR}$ given by Eq. (\ref{fR3r}). All the figures  are plotted using the following values of the constants, $M=1$, $c_0=0.01$, $c_1=0.001$, $c_2=-10^{5}$, $c_3=1$ and $c_4=1$. These values satisfy the constrains given by Eq. (\ref{cond}). }}
\label{Fig:3}
\end{figure}

The asymptote form of the temporal component of the BH solution (\ref{sp1}) has the form:
\begin{equation}\label{hor}
g_{t\,t}\approx \frac{24{r}^{2}c_0{}^3\ln\,c_0}{c_1} -{r}^{2}c_2-3c_0{}^2-\frac {6c_0{c_1}}{{r}^{2}}\,.
\end{equation}
Equation (\ref{hor}) has four real roots given as:

\begin{eqnarray}\label{hor1}
&&r_{_{\pm}}=\frac { \sqrt {\left(\pm\sqrt{9\,{c_0}^{4}+576\,{c_0}^{4}\ln  c_0-24\,c_0\,c_2\,c_1}- 3\,{c_0}^{2} \right) c_1}}{\sqrt{2(24\,{
c_0}^{3}\ln \,c_0-c_2\,c_1)}}
\nonumber\\
 &&r_{_{(1,2)}}=-\frac { \sqrt {\left(\pm\sqrt{9\,{c_0}^{4}+576\,{c_0}^{4}\ln  c_0-24\,c_0\,c_2\,c_1}- 3\,{c_0}^{2} \right) c_1}}{\sqrt{2(24\,{
c_0}^{3}\ln \,c_0-c_2\,c_1)}}
\,.\nonumber\\
 &&
\end{eqnarray}
where $r_{_{(\pm)}}$ are the inner and outer horizons of the spacetime.  In Fig.~\ref{Fig:3} \subref{fig:metpot} we plot the metric potentials of $g_{t\,t}$ and $g_{r\,r}$ showing their behavior.  In Fig.~\ref{Fig:3} \subref{fig:hor} we show the horizons given by Eq. (\ref{hor1}) showing that for the specific values of the constant $c_2$ we can have two horizons, inner and out, or the two horizons coincide constitute a degenerate horizon or we can enter a region where there is no horizon, appearance  naked singularity.

The  temperature of Hawking is given by \cite{PhysRevD.86.024013,Sheykhi:2010zz,Hendi:2010gq,PhysRevD.81.084040,Wang:2018xhw,Zakria:2018gsf}
\begin{eqnarray}\label{temp}
&&T(r_+) = \frac{g'_{tt}}{4\pi}=\frac {24\,{r}^{6}\ln  \left( c_0 +\frac{c_1}{{r}^{2}}
 \right) {c_0}^{3}[r^2c_0+c_1]-{r}^{8}c_2\,c_1,c_0-{r}^{6}c_2\,{c_1}^{2}-24\,{r}^{6}{c_0}^{3}c_1
-6{r}^{4}{c_0}^{2}{c_1}^{2}-4{r}^{2}c_0{c_1}^{3}+2{c_1}^{4}}{
\pi \, \left( c_0{r}^{2}+c_1\right) c_1{r}^{5}}\,.\nonumber\\
&&
\end{eqnarray}
In Fig.~\ref{Fig:3} \subref{fig:temp} we show the behavior of the Hawking temperature of the BH solution (\ref{sp1}) showing that the temperature is always positive.
The semi classical Bekenstein-Hawking entropy of the horizons is defined as
\begin{equation}\label{ent}
S(r_+)=\frac{{\cal A}}{4G} f_R(r_+)=4\pi^2r_1f_R(r_+)=4\pi^2r_+\left(c_0+\frac{c_1}{r_+{}^2}\right)\,,
\end{equation}
with ${\cal A}=2\pi r_+$  being the area of the event horizons and
the gravitational constant $G$ equals $G=\frac{1}{8\pi}$.  The behavior  of the entropy is depicted in Fig.~\ref{Fig:3} \subref{fig:ent1}  showing that the BH solution given by Eq. (\ref{sp1}) has always positive entropy.
Finally,   the heat capacity   is figured out as \cite{Zheng:2018fyn,Kim:2012cma}
\begin{eqnarray} \label{heat3}
&&C(r_+)=T(r_+)\left(\frac{S'(r_+)}{T'(r_+)}\right)=\Bigg\{ \Bigg( 24\,{r}^{8}\ln  \left( c_0\,{r}^{2}+c_1
 \right) {c_0}^{4}+24\,{r}^{6}\ln  \left( c_0\,{r}^{2}+c_1 \right) {c_0}^{3}c_1-{r}^{8}c_2\,c_1\,c_0-{r}^{6}c_2\,{c_1}^{2}-24\,{r}^{6}{c_0}^{3}c_1\nonumber\\
 &&-48\,{r}^{8}\ln \,r {c_0}^{4}-48\,{r}^{6}\ln
 \,r {c_0}^{3}c_1-6\,{r}^{4}{c_0}^{2}{c_1}^{2}-4\,{r}^{2}c_0\,{c_1}^{3}+2\,{c_1}^{4}
 \Bigg) {\pi }^{2} \left( {r}^{4}{c_0}^{2}-{c_1}^{2}
 \right)\Bigg\}\Bigg\{4r \Bigg( 24\,{r}^{10}\ln  \left( c_0\,{r}^{2}+c_1
 \right) {c_0}^{5}\nonumber\\
 &&-24\,{r}^{8}{c_0}^{4}c_1-{r}^{6}c_2\,{c_1}^{3}-54\,{r}^{6}{c_0}^{3}{c_1}^{2}-48\,{r}^
{10}\ln \,r {c_0}^{5}+26\,{r}^{4}{c_0}^{2}{c_1}^{3}-2\,{r}^{2}c_0\,{c_1}^{4}+48\,{r}^{8}\ln
 \left( c_0\,{r}^{2}+c_1 \right) {c_0}^{4}c_1\nonumber\\
 &&+24\,
{r}^{6}\ln  \left( c_0\,{r}^{2}+c_1 \right) {c_0}^{3}{c_1}^{2}-{r}^{10}c_2\,c_1\,{c_0}^{2}-2\,{r}^{8}c_2\,{c_1}^{2}c_0-96\,{r}^{8}\ln\,r {c_0}^{4}c_1-48\,{r}^{6}\ln \,r {c_0}^{3}{c_1}^{2}-10\,{c_1}^{5} \Bigg)\Bigg\}^{-1}
\,,
\end{eqnarray}
where $S'(r_1)$ and $T'(r_1)$ are the derivative of entropy and Hawking temperature with respect to the outer horizon respectively.  We depict the behavior of the heat capacity given by Eq. (\ref{heat3})   in Fig.~\ref{Fig:3} \subref{fig:heat1} which shows that we have a stable model because the heat capacity is always positive.
\subsection{Thermodynamics of the BH  (\ref{sp3})}\label{S42}
\begin{figure}
\centering
\subfigure[~The behavior of the metric potential given by Eq. (\ref{sp3})]{\label{fig:metpotr}\includegraphics[scale=0.23]{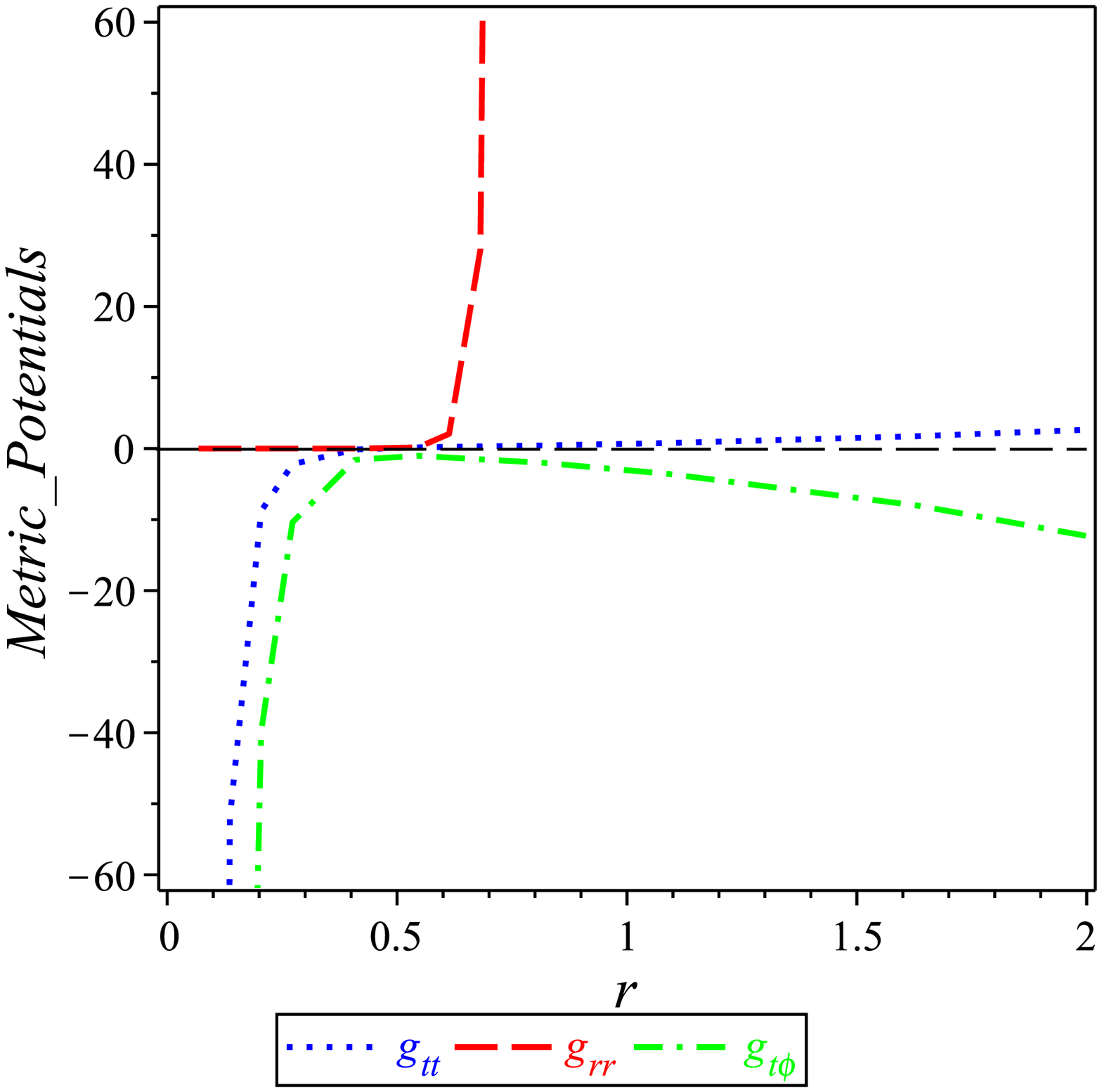}}
\subfigure[~The behavior of the horizons  given   by Eq.  (\ref{hor33})]{\label{fig:horr}\includegraphics[scale=0.23]{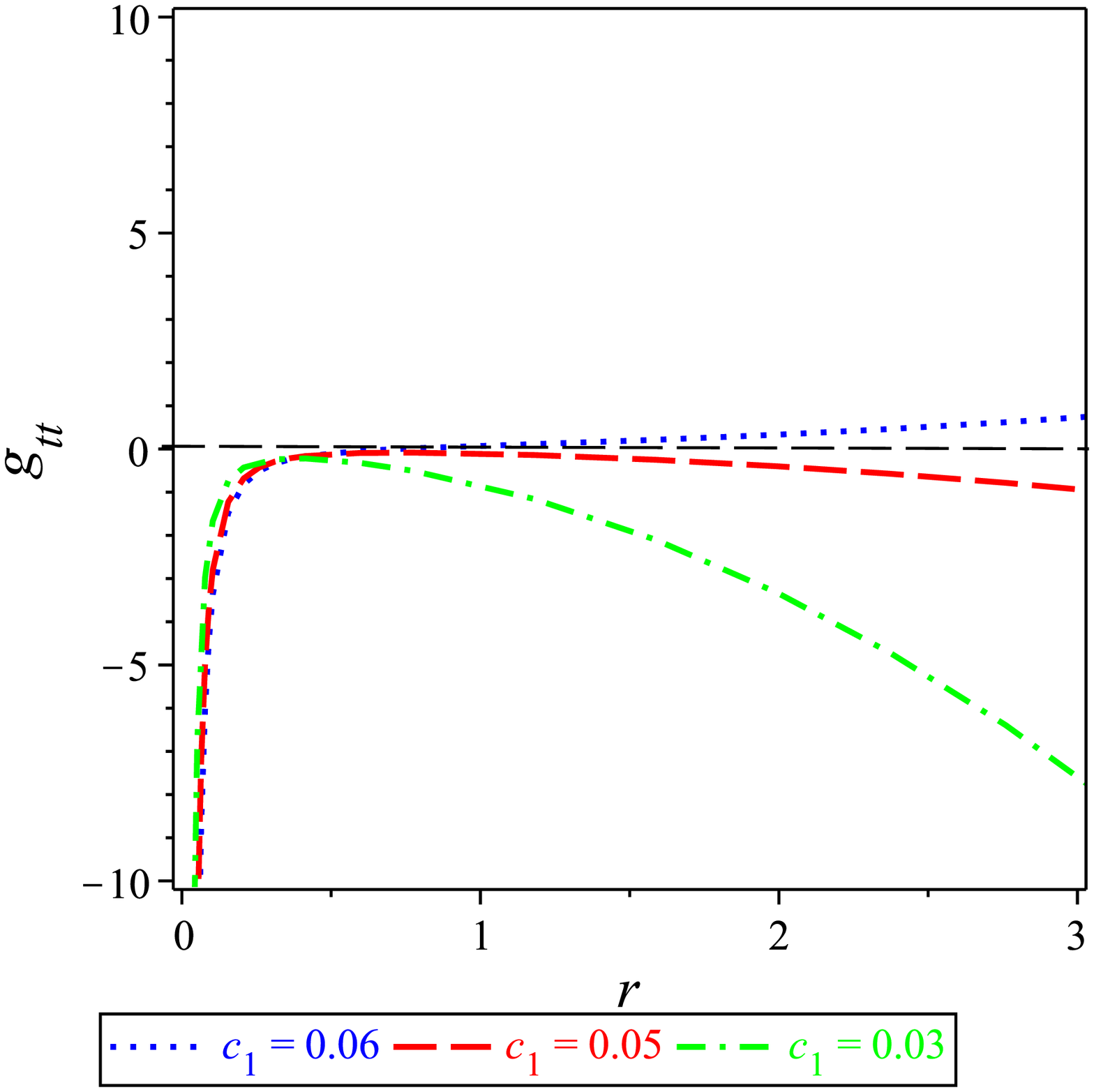}}
\subfigure[~The behavior of the Hawking temperature  given   (\ref{temp3})]{\label{fig:tempr}\includegraphics[scale=0.23]{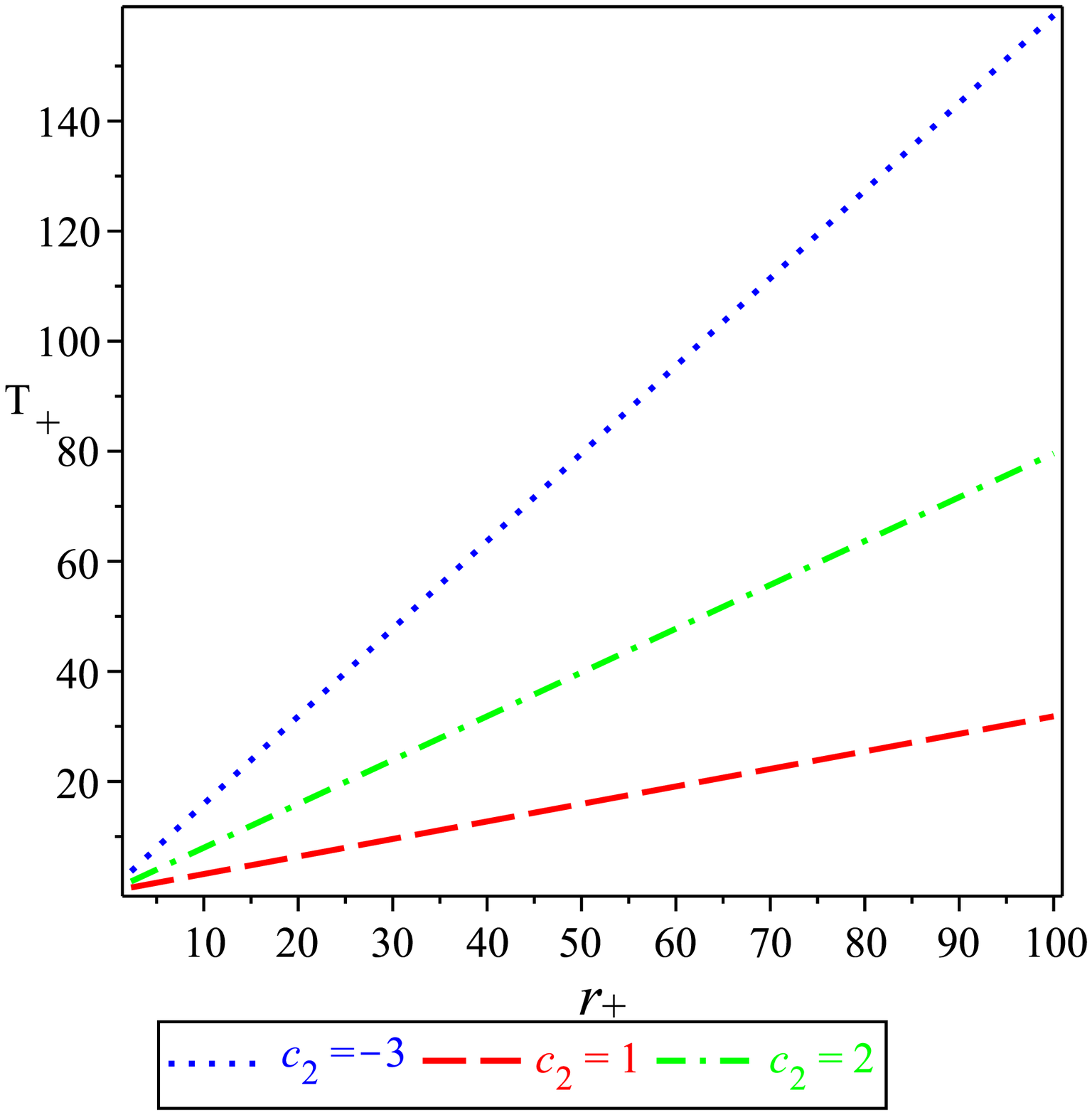}}
\subfigure[~The behavior of the heat capacity  given   (\ref{heat3})]{\label{fig:heat1r}\includegraphics[scale=0.23]{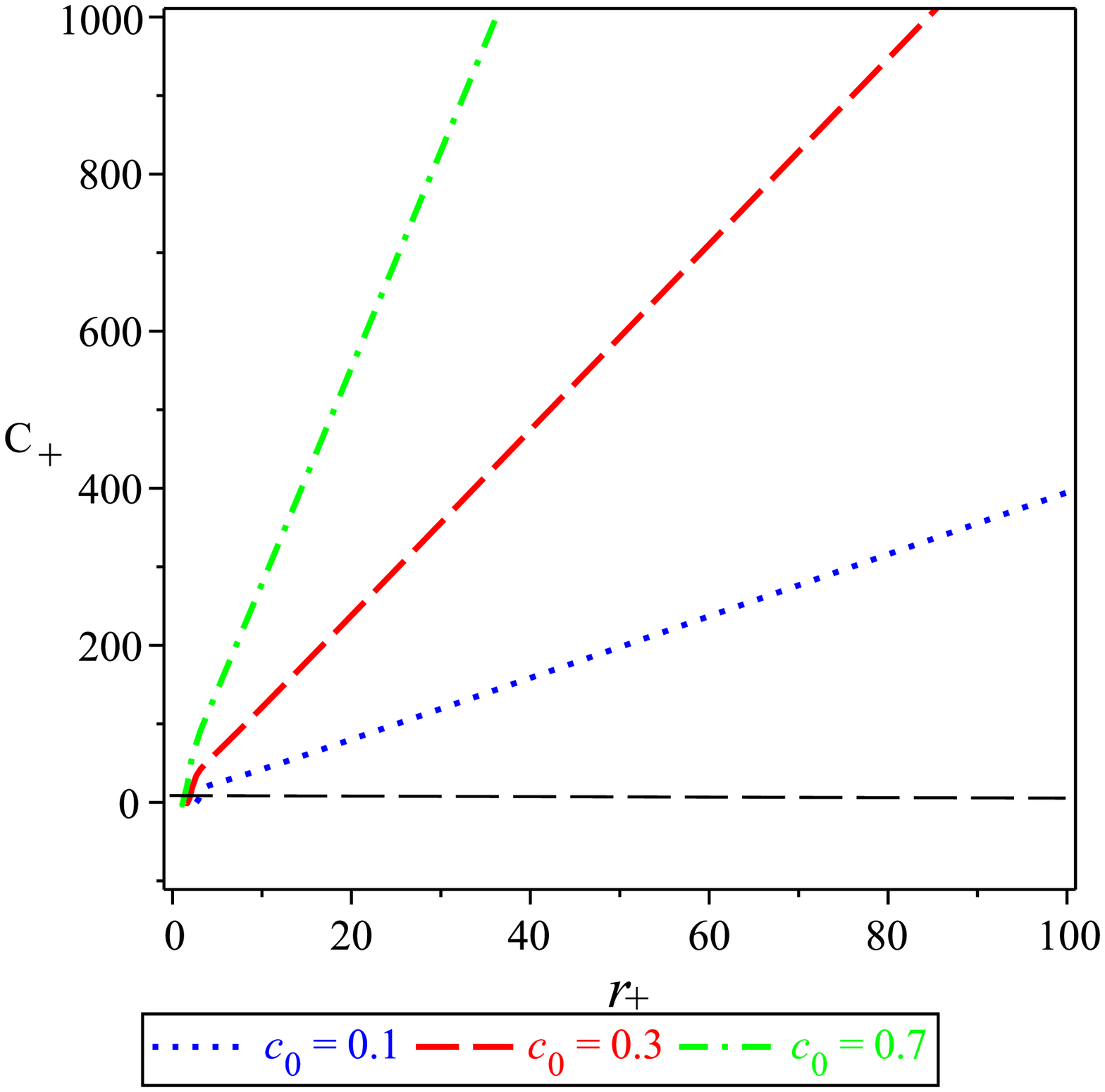}}
\caption[figtopcap]{\small{Systematic plots of; \subref{fig:metpotr} the metric potentials  given by Eq. (\ref{sp3});   \subref{fig:horr} the horizons of the temporal component of $g_{tt}(r)$ given   (\ref{hor3});  \subref{fig:tempr} the Hawking temperature given by Eq. (\ref{temp3}), and  \subref{fig:heat1r} the heat capacity given by Eq. (\ref{heat3}). All the figures  are plotted using the following values of the constants, $M=1$, $c_0=0.01$, $c_1=0.001$, $c_2=-10^{5}$, $c_3=1$ and $c_4=1$. These values satisfy the constrains given by Eq. (\ref{cond}). }}
\label{Fig:4}
\end{figure}

Now we are going the same steps used in the non-rotating case and get the asymptote form of the temporal component of the BH solution (\ref{sp3}) to has the following form:
\begin{equation}\label{hor2}
g_{t\,t}\approx \frac {24\,{r}^{2}{c_0}^{3}\ln c_0 }{c_1}-{r}^{2}c_2+{r}^{2}{c_4}^{2}-48\,{r}^{2}{c_0}^
{3}c_4\,\ln\,c_0 +3\,{c_0}^{2}-6\,c_1\,{c_0}^{2}c_4+12\,{\frac {{c_1}^{2}c_4
\,c_0}{{r}^{2}}}-6\,{\frac {c_1\,c_0}{{r}^{2}}}\,.
\end{equation}
Equation (\ref{hor2}) has four real roots given as:
\begin{eqnarray}\label{hor3}
&&r_{_{(1,2)}}=\pm\frac {\sqrt [4]{c_0c_1{}^2}\sqrt {3\,{c_0}^{3/2}[2c_4\,c_1-1]+\sqrt {3}\sqrt {2\,c_1\,c_4-1}\sqrt {6\,{c_0}^{3}c_4\,c_1-8\,{c_4}^{2}c_1 -3\,{c_0}^{3}+192\,{c_0}^{3}\ln  c_0[2c_4\,c_1-1] +8\,c_2\,c_1}}}{\sqrt {2(24 \,{c_0}^{3}\ln c_0-c_2\,c_1+{c_4}^{2}c_1-48\,{c_0}^{3}c_4\,c_1\,\ln c_0 )}}\,,
\nonumber\\
 &&r_{_{(5,6)}}=\pm\frac {\sqrt [4]{c_0c_1{}^2}\sqrt {3\,{c_0}^{3/2}[2c_4\,c_1-1]-\sqrt {3}\sqrt {2\,c_1\,c_4-1}\sqrt {6\,{c_0}^{3}c_4\,c_1-8\,{c_4}^{2}c_1 -3\,{c_0}^{3}+192\,{c_0}^{3}\ln  c_0[2c_4\,c_1-1] +8\,c_2\,c_1}}}{\sqrt {2(24
\,{c_0}^{3}\ln c_0-c_2\,c_1+{c_4}^{2}c_1-48\,{c_0}^{3}c_4\,c_1\,\ln c_0 )}}\,.\nonumber\\
 &&
\end{eqnarray}
  As Eq. (\ref{hor3}) shows that there will  real roots if  $c_0>0$, $c_1>\frac{1}{c_4}$  and\\ $6\,{c_0}^{3}c_4\,c_1-8\,{c_4}^{2}c_1 -3\,{c_0}^{3}+192\,{c_0}^{3}\ln  c_0[2c_4\,c_1-1] +8\,c_2\,c_1>0$. In this case we will deal with one horizon which we call it $r_h$ that given by
  \begin{eqnarray}\label{hor33}
&&r_{_h}=\frac {\sqrt [4]{c_0c_1{}^2}\sqrt {3\,{c_0}^{3/2}[2c_4\,c_1-1]+\sqrt {3}\sqrt {2\,c_1\,c_4-1}\sqrt {6\,{c_0}^{3}c_4\,c_1-8\,{c_4}^{2}c_1 -3\,{c_0}^{3}+192\,{c_0}^{3}\ln  c_0[2c_4\,c_1-1] +8\,c_2\,c_1}}}{\sqrt {2(24
\,{c_0}^{3}\ln c_0-c_2\,c_1+{c_4}^{2}c_1-48\,{c_0}^{3}c_4\,c_1\,\ln c_0 )}}\,.\nonumber\\
&&
\end{eqnarray}
  In Fig.~\ref{Fig:4} \subref{fig:metpotr} we plot the metric potentials of $g_{t\,t}$, $g_{r\,r}$, and $g_{t\,\phi}$ showing their behavior.  In Fig.~\ref{Fig:4} \subref{fig:horr} we show the horizons given by Eq. (\ref{hor33}) showing that for  specific values of the constant $c_1$ we can have one horizon a degenerate horizon or we can enter a region where there is no horizon, appearance  naked singularity.

The  temperature of Hawking of the BH solution (\ref{sp3}) is given by \cite{PhysRevD.86.024013,Sheykhi:2010zz,Hendi:2010gq,PhysRevD.81.084040,Wang:2018xhw,Zakria:2018gsf}:
\begin{eqnarray}\label{temp3}
&&T(r_{_h}) =\frac {1}{2\pi \, \left( c_0\,{r}^{2}+c_1 \right) {r} ^{5}c_1}\Bigg\{24\,{r}^{6}{c_0}^{3}\ln  \left( c_0\,{r}^{2}+c_1 \right) c_1-48\,{r}^{6}{c_0}^{3}\ln\,r c_1+48\,{r}^{6}{c_0}^{3}{c_1}^{2}c_4+12 \,{r}^{4}{c_1}^{3}{c_0}^{2}c_4+8\,{r}^{2}c_0\,{c_1}^{4}c_4\nonumber\\
 &&-{r}^{8}c_1\,c_2\,c_0+{r}^{8}c_1\,{c_4}^{2}c_0-4\,{c_1}^{5}c_4+24\,{r}^ {8}{c_0}^{4}\ln  \left( c_0\,{r}^{2}+c_1 \right) -48\,{ r}^{8}{c_0}^{4}\ln\,r -24\,{r}^{6}{c_0}^{3}c_1-6\,{r}^{4}{c_0}^{2}{c_1}^{2}-4\,{r}^{2}c_0\,{c_1}^{3}-{r}^{6}c_2\,{c_1}^{2}\nonumber\\
 &&+{r}^{6}{c_4}^{2} {c_1}^{2}-48\,{r}^{8}{c_0}^{4}\ln  \left( c_0\,{r}^{2}+ c_1 \right) c_1\,c_4-48\,{r}^{6}{c_0}^{3}\ln \left( c_0\,{r}^{2}+c_1 \right) {c_1}^{2}c_4+96 \,{r}^{8}{c_0}^{4}\ln\,r c_1\,c_4+96\, {r}^{6}{c_0}^{3}\ln  \,r {c_1}^{2}c_4+2 \,{c_1}^{4}\Bigg\}\,.\nonumber\\
 &&
\end{eqnarray}
In Fig.~\ref{Fig:4} \subref{fig:tempr} we show the behavior of the Hawking temperature of the BH solution (\ref{sp3}) showing that the temperature is always positive.
Finally,   the heat capacity   of the BH solution (\ref{sp3}) is figured out as \cite{Zheng:2018fyn,Kim:2012cma}
\begin{eqnarray} \label{heat}
  &&C(r_{_h})=T(r_{_h})\left(\frac{S'(r_{_h})}{T'(r_{_h})}\right)=\Bigg[4\Bigg( 2\,{c_1}^{4}+24\,{r}^{6}{c_0}^{3}\ln \left( c_0\,{r}^{2}+c_1 \right) c_1-48\,{r}^{6}{c_0}^{3}\ln\,r c_1+48\,{r}^{6}{c_0}^{3}{c_1}^{2}c_4+12\,{r}^{4}{c_1}^{3}{c_0}^{2}c_4\nonumber\\
 &&+8 \,{r}^{2}c_0\,{c_1}^{4}c_4-{r}^{8}c_1\,c_2\,c_0+{r}^{8}c_1\,{c_4}^{2}c_0-48\,{r}^{8}{c_0}^{4}\ln  \left( c_0\,{r}^{2}+c_1 \right) c_1\,c_4-48\,{r}^{6}{c_0}^{3}\ln  \left( c_0\,{r}^{2}+c_1 \right) {c_1}^{2}c_4+96\,{r}^{8}{c_0}^{4}\ln\,r c_1\,c_4\nonumber\\
 &&+96\,{r}^{6}{c_0}^{3}\ln \,r {c_1}^{2}c_4-4\,{c_1}^{5}c_4 +24\,{r}^{8}{c_0}^{4}\ln  \left( c_0\,{r}^{2}+c_1 \right) -48\,{r}^{8}{c_0}^{4}\ln\,r -24\,{r}^{6}{ c_0}^{3}c_1-6\,{r}^{4}{c_0}^{2}{c_1}^{2}-4\,{r}^{2 }c_0\,{c_1}^{3}-{r}^{6}c_2\,{c_1}^{2}\nonumber\\
 &&+{r}^{6}{c_4}^{2}{c_1}^{2} \Bigg) {\pi }^{2} \left( {c_0}^{2}{ r}^{4}-{c_1}^{2} \right)\Bigg]\Bigg\{r \Bigg( 108\,{r}^{6}{c_1}^{3}{c_0}^{3}c_4+48\,{r}^{8}{c_0}^{4}{c_1}^{2}c_4-{r}^{10}c_1\,c_2\,{c_0}^{2}-52\,{r}^{4}{c_1}^{4}{c_0}^{2}c_4-2\,{r}^{8}{c_1}^{2}c_2\,c_0+{r}^{10}c_1\,{c_4}^{2}{c_0}^{2}\nonumber\\
 &&+2\,{r}^{8}{c_1}^{2}{c_4}^{2}c_0+4\,{r}^{2}c_0\,{c_1}^{5 }c_4-48\,{r}^{10}{c_0}^{5}\ln  \left( c_0\,{r}^{2}+c_1 \right) c_1\,c_4-96\,{r}^{8}{c_0}^{4}\ln \left( c_0\,{r}^{2}+c_1 \right) {c_1}^{2}c_4-48 \,{r}^{6}{c_0}^{3}\ln  \left( c_0\,{r}^{2}+c_1 \right) {c_1}^{3}c_4\nonumber\\
 &&+96\,{r}^{10}{c_0}^{5}\ln\,r c_1\,c_4+192\,{r}^{8}{c_0}^{4}\ln\,r {c_1}^{2}c_4+96\,{r}^{6}{c_0}^{3}\ln\,r {c_1}^{3}c_4+20\,{c_1}^{6}c_4+26\,{r}^{4}{c_1}^{3}{c_0}^{2}-2\,c_0\,{r}^{2}{c_1}^{4}-54\,{r}^{6}{c_0}^{3}{c_1}^{2}\nonumber\\
 &&+48\,{r}^{8}{c_0}^{ 4}\ln  \left( c_0\,{r}^{2}+c_1 \right) c_1+24\,{c_0}^{3}\ln  \left( c_0\,{r}^{2}+c_1 \right) {r}^{6}{c_1} ^{2}-96\,{r}^{8}{c_0}^{4}\ln\,r c_1-48\,{c_0}^{3}\ln\,r {r}^{6}{c_1}^{2}-10\,{c_1} ^{5}-24\,{r}^{8}{c_0}^{4}c_1+{r}^{6}{c_4}^{2}{c_1}^{3}\nonumber\\
 &&-{r}^{6}c_2\,{c_1}^{3}+24\,{r}^{10}{c_0}^{5} \ln  \left( c_0\,{r}^{2}+c_1 \right) -48\,{r}^{10}{c_0} ^{5}\ln\,r  \Bigg) \Bigg\}^{-1}\,,
\end{eqnarray}
where $S'(r_{_h})$ and $T'(r_{_h})$ are the derivative of entropy and Hawking temperature concerning the event horizon respectively.  We depict the behavior of the heat capacity given by Eq. (\ref{heat})   in Fig.~\ref{Fig:4} \subref{fig:heat1r} which shows that we have a stable model because the heat capacity is always positive.

\section{ Discussion and conclusions }\label{S77}
In this work, we have explored two new classes of three
dimensional BH solutions in $f(R)$ gravity. By varying the action
we derived the field equations. Taking the trace of the field
equations, we have rewritten the field equations in the form of
Eq. (5). We applied the form of the field equation of $f(R)$,
written in terms of $f_R$, to 3D spacetime that has three unknown
functions, one of them is responsible for making the metric to
have a rotating form, i.e., $b_2(r)$. We classified the resulting
field equations into four cases: (i) $f_R=constant$ and
$b_2(r)=0$, (ii)$f_R=constant$ and $b_2(r)\neq0$, (iii) $f_R\neq
constant$ and $b_2(r)=0$, and finally, (iv) $f_R\neq constant$ and
$b_2(r)\neq 0$. We focus on the last two cases because the first
two coincide with the rotating/non-rotating three dimensional
solutions of GR.

Assuming the form of $f_R(r)=c_0+\frac{c_1}{r^2}$ we are able to
solve the field equation with/without an  unknown rotating
function. Our solutions cannot coincide with the 3D solutions of
GR because the constant $c_1$  is not equal to zero. Although our
field equation do not involve the cosmological constant term,
however the study of the asymptote of these BH solutions showed
that they behave as AdS/dS. In fact, the constant $c_0$ plays the
role of the cosmological constant which indeed means that this
constant cannot be vanished. We also showed that the invariants of
these BHs have a true singularity at the origin and a strong one
as compared to the 3D solutions of GR. The source of the strong
singularity comes from the non-trivial forms of the Ricci scalar
of the two BHs. We calculated the form of $f(R)$ of the BHs and
showed that both of them behave as a polynomial one. Moreover, we
calculated the second derivative of $f(R)$, i.e., $f_{RR}$ of the
two BHs and showed analytically and graphically that they have
positive values,   which means that the Dolgov-Kawasaki stability
criterion is satisfied \cite{Hendi:2014mba} and this ensures that
our BHs avoid tachyonic instabilities
\cite{DeFelice:2010aj,Wang:2012rw,Faraoni:2006sy,Cognola:2007zu}.

We also investigated the causal structure of the solutions and
showed that they possess several horizons. We found out that these
solutions show stable thermodynamic behavior in all regions since
the Hawking temperature and heat capacity are always positive and
free from singular points, which indeed ensure stable BHs
thermodynamics configuration where no phase transitions occur.

Finally, we would like to mention that many issues remain for
further investigations. One may consider the geodesic structure
and Penrose diagrams of these spacetimes. Besides, the techniques
presented in \cite{Oliveira-Neto:1996szd,Tavlayan:2020chf} could
be applied for describing the issue of the structure and nature of
the horizons and singularities of the obtained solutions. It would
be also interesting to explore thermodynamic properties of the BH
solutions, as well as possible holographic applications.


\end{document}